\documentclass[aip,apl,amsmath,amssymb,reprint, groupedaddress]{revtex4-1}

\usepackage{graphicx}% Include figure files
\usepackage{dcolumn}% Align table columns on decimal point
\usepackage{bm}% bold math

\usepackage[utf8]{inputenc}
\usepackage[T1]{fontenc}
\usepackage{mathptmx}
\usepackage{etoolbox}

\usepackage{siunitx} % \SI{22}{\micro\ampere}, etc.
\usepackage[colorlinks=true,linkcolor=blue,citecolor=blue,urlcolor=black]{hyperref}
\graphicspath{{figures/}} % dossier des figures

\makeatletter
\def\@email#1#2{%
 \endgroup
 \patchcmd{\titleblock@produce}
  {\frontmatter@RRAPformat}
  {\frontmatter@RRAPformat{\produce@RRAP{*#1\href{mailto:#2}{#2}}}\frontmatter@RRAPformat}
  {}{}
}%
\makeatother
\begin{document}

%\preprint{AIP/123-QED}

\title{Reconfigurable Superconducting Logic for On-Chip Photon Coincidence Detection}

% Force line breaks with \\
\author{Gabriel Le Guay}
% \email{gablg@mit.edu}
\thanks{These authors contributed equally to this work. \\ Corresponding authors: \href{mailto:gablg@mit.edu}{\textcolor{blue}{gablg@mit.edu}}, \href{mailto:mcaste@mit.edu}{\textcolor{blue}{mcaste@mit.edu}}}

\author{Matteo Castellani}
\thanks{These authors contributed equally to this work. \\ Corresponding authors: \href{mailto:gablg@mit.edu}{\textcolor{blue}{gablg@mit.edu}}, \href{mailto:mcaste@mit.edu}{\textcolor{blue}{mcaste@mit.edu}}}

\affiliation{%
Research Laboratory of Electronics,\\
Massachusetts Institute of Technology,\\
Cambridge, Massachusetts 02139, United States
}%

\author{Reed Foster}

\affiliation{%
Research Laboratory of Electronics,\\
Massachusetts Institute of Technology,\\
Cambridge, Massachusetts 02139, United States
}%

\author{Francesca Incalza}

\affiliation{%
Research Laboratory of Electronics,\\
Massachusetts Institute of Technology,\\
Cambridge, Massachusetts 02139, United States
}%

\author{Alejandro Simon}

\affiliation{%
Research Laboratory of Electronics,\\
Massachusetts Institute of Technology,\\
Cambridge, Massachusetts 02139, United States
}%

\author{Owen Medeiros}

\affiliation{%
Research Laboratory of Electronics,\\
Massachusetts Institute of Technology,\\
Cambridge, Massachusetts 02139, United States
}%
\author{Phillip D. Keathley}

\affiliation{%
Research Laboratory of Electronics,\\
Massachusetts Institute of Technology,\\
Cambridge, Massachusetts 02139, United States
}%

\author{Karl K. Berggren}

\affiliation{%
Research Laboratory of Electronics,\\
Massachusetts Institute of Technology,\\
Cambridge, Massachusetts 02139, United States
}%

\begin{abstract}

Scaling photonic quantum-information platforms requires arrays of superconducting nanowire single-photon detectors (SNSPDs) for feedforward control, in which optical operations are conditioned on Bell-state measurements relying on photon-coincidence detections. On-chip superconducting cryotron electronics, performing logic on detector outputs and driving optical modulators, could reduce latency and room-temperature interconnect complexity for feedforward schemes. To date, no cryotron circuits designed for this purpose have been demonstrated. We demonstrate a bias-programmable logic gate based on three nanocryotrons (nTrons) that implements selectable AND (coincidence), XOR (odd-parity), and OR functions. It operates on two electrical pulses at 4.2\,K, with bit-error rates below $10^{-3}$, bias margins up to $\pm21.7\%$, and operation extending to 25\,MHz over narrower bias windows. It performs coincidence and odd-parity detection on two SNSPDs’ outputs with bit-error rates below $3.2 \times 10^{-2}$. As proof-of-concept, we show that nTrons can drive capacitive loads up to 1.15\,V, potentially enabling compatibility with electro-optic modulators in feedforward schemes.

\end{abstract}

\maketitle

Large-scale arrays of superconducting nanowire single-photon detectors (SNSPDs) have the potential to significantly advance quantum communication \cite{Wang2020Integrated} and
quantum computing \cite{kwon2025manufacturable}. In particular, arrays of waveguide-integrated detectors \cite{Colangelo:20, AlSayem:20} will help scale up quantum photonic platforms substantially, by enabling on-chip Bell-state measurements, a fundamental operation in quantum protocols \cite{kwon2025manufacturable, sinclair_spectral_2014}.
Typically, these protocols also require feedforward control, in which photon-detection outcomes rapidly influence subsequent optical components through classical computation \cite{sinclair_spectral_2014}. Reading out arrays and performing such classical computation with room-temperature electronics requires several cryostat feedthroughs, limiting scalability and increasing latency. 
Multiplexing schemes \cite{doerner_frequency-multiplexed_2017, wollman_kilopixel_2019, oripov_superconducting_2023} and cryogenic processors based on cryo-CMOS \cite{viskova_cryo-cmos_2022, fredenburg_32-channel_2024} or rapid single-flux-quantum (RSFQ) \cite{Miyajima2018SFQEncoder, Yabuno2020Scalable} have been explored to reduce wiring in SNSPD arrays, and CMOS discrete logic elements have been used to implement feedforward control with single detectors at cryogenic temperatures \cite{Thiele:25, lamberty_interfacing_2025}. However, these approaches require additional fabrication processes, increase power dissipation, or show limited compatibility with the high impedance of nanowires, modulators, and CMOS technologies. An appealing alternative is to perform signal processing directly in the nanowire platform itself using nanocryotrons. 

Nanocryotrons (nTrons) are three-terminal nanowire devices fabricated on a single thin-film superconducting layer in which an electrically coupled gate controls the switching current of the channel \cite{nTron}. 
Nanocryotrons share the fabrication process of SNSPDs, are robust to millitesla magnetic fields\cite{Buzzi2024Nanocryotron}, dissipate as little as 20\,aJ/pulse \cite{Simon2025CryotronGeometry}, can amplify small signals and drive high impedances \cite{Zhao2017nTronComparator, paul_photolithography-compatible_2025}, making them promising candidates for integrated cryogenic electronics. Their functionalities arise from engineering the device geometry, which in principle enables their implementation in a variety of superconducting materials used for waveguide-integrated SNSPDs \cite{kwon2025manufacturable, Colangelo2024Molybdenum, buckley_all-silicon_2017}. 
Previous demonstrations in niobium nitride (NbN) and niobium titanium nitride (NbTiN) have shown counters \cite{Castellani2024Counter} and encoders \cite{Huang2024Encoder, Zheng2020BinaryEncoder} interfaced with SNSPDs, shift-registers \cite{Foster2023ShiftRegister}, memories \cite{butters_scalable_2021, medeiros_scalable_2026}, logic gates \cite{nTron,Buzzi2024Nanocryotron, Wang2025Attojoule}, rectifiers \cite{Castellani2025Rectifier}, and neuromorphic components \cite{Toomey2020SNWSpiking, lombo_superconducting_2022}.

\begin{figure*}[t]
    \centering
    \includegraphics[width=0.95\linewidth]{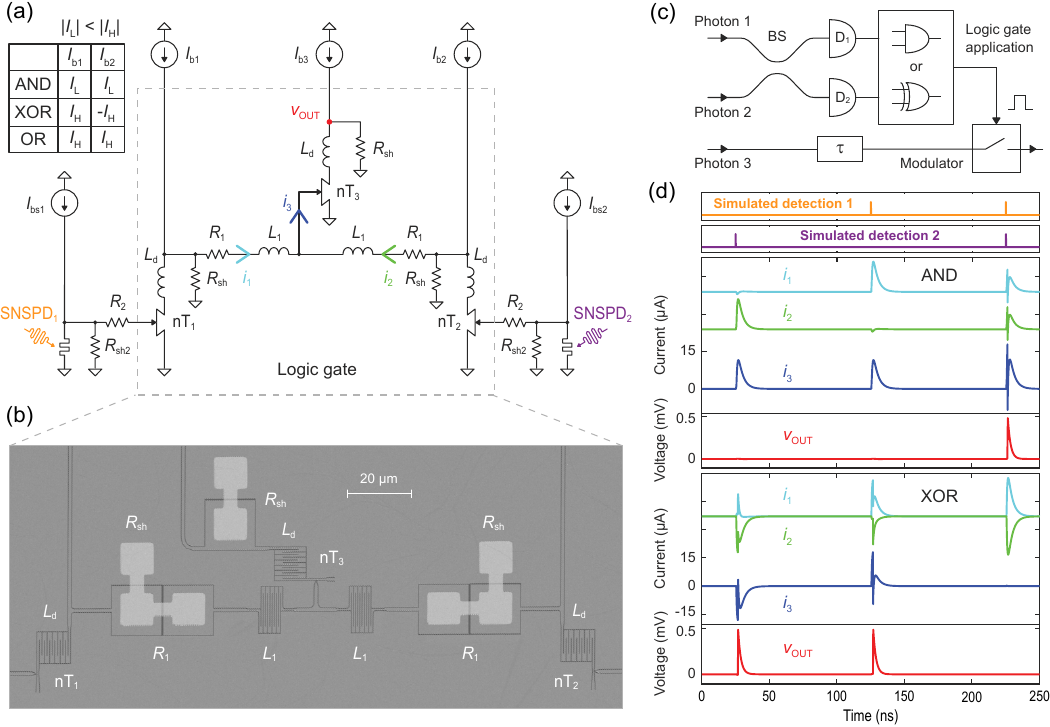} 
    \caption{Reconfigurable logic gate for single-photon coincidence detection with simulations, fabrication, and application. (a) Circuit schematic of the gate interfaced with two SNSPDs (SNSPD$_1$ and SNSPD$_2$). The two side nTrons (nT$_1$ and nT$_2$) receive independent input pulses from the detectors, while the central nTron (nT$_3$) produces the output. The bias currents \(I_\text{b1}\) and \(I_\text{b2}\) set the operating configuration (see the table, top left). Parameters: $L_1 = 10\,\mathrm{nH}$, $L_\text{d} = 8\,\mathrm{nH}$, $R_1 = 10\,\Omega$, $R_2 = 5\,\Omega$, $R_{\mathrm{sh}} = 5\,\Omega$, $R_{\mathrm{sh2}} = 50\,\Omega$.
    (b) Scanning electron micrograph (SEM) of the fabricated three-nTron circuit with bias and output traces. White labels associate the components with the circuit elements in (a). The Ti/Au resistors are in light gray, the gray regions correspond to the niobium nitride (NbN) film, and the dark edges of the traces reveal the underlying SiO\(_2\) substrate. A continuous ground plane surrounds all the circuit traces.
    (c) Schematic of the envisioned gate application in a photonic feedforward setup. Two photons enter a directional coupler (beam splitter, BS) and exit either together through the same port, due to interference, or separately through different ports. Detection events at D$_1$ and D$_2$ (SNSPD$_1$ and SNSPD$_2$) are processed by the reconfigurable logic gate in AND or XOR mode, and the gate output pulses subsequently control an optical switch acting on a photon that propagates through a different waveguide. An optical delay line is incorporated into the waveguide to synchronize photon arrival with the circuit processing latency.
    (d) Time-domain simulation of the gate dynamics for the XOR and AND configurations. From top to bottom, the panels show the unitless control signals used to generate photon detection events in the detectors, the corresponding internal currents \(i_1\) and \(i_2\), their sum \(i_3\), and the output voltage \(v_{\mathrm{OUT}}\). For visual clarity and compactness, the traces \(i_1\) and \(i_2\) are vertically offset in both the XOR and AND panels. Bias currents: $I_\text{bs1}=I_\text{bs2}=25$\,\textmu A, $I_\text{b3}=112$\,\textmu A, $I_\text{L}=90$\,\textmu A, $I_\text{H}=110$\,\textmu A.}
    \label{fig:1}
\end{figure*}

Despite the wide variety of circuits proposed to date, no cryotron logic gates specifically optimized to process SNSPD outputs for quantum applications have been demonstrated. In a typical beam-splitter interferometric setup of Hong–Ou–Mandel (HOM) experiments and linear-optical Bell-state measurements\cite{kwon2025manufacturable}, a reconfigurable cryotron gate integrated with two SNSPDs would enable the selective detection of either photon coincidence or bunching events. 
More broadly, a single circuit capable of implementing a universal set of logic functions would enable more complex operations, including the real-time classical processing required in photonic quantum protocols, with limited design overhead.
In the cryotron gates demonstrated in prior works, certain logic functions required dedicated circuit designs that limit flexibility after fabrication, or device reset always relied on external signals that potentially add power overhead and design complexity when only combinational operations are needed. 
Alam~et~al. proposed an approach to improve design flexibility by realizing multiple basic logic functions (AND, OR, NOT, MAJ) within a single cryotron \cite{Alam2024HeaterCryotron}. Their work exploits the addition of input currents and reconfigurable biasing conditions to achieve a universal gate, with an external reset signal. 

In this work, we present a fully reconfigurable nanocryotron logic gate that can perform three operations: AND, OR, and XOR, within the same circuit without an external reset signal. This work builds on the concepts of Alam~et~al. but introduces additional nanocryotrons to amplify detector signals and directly implement XOR operations, which facilitates photon bunching detection. Moreover, it operates using short current pulses comparable to those produced by SNSPDs, and it self-resets after each operation thanks to on-chip shunt resistors.
Adding an external pulsed signal to one input of an XOR gate implements the NOT function, enabling a universal combinational logic family using a single device.
We demonstrate XOR and AND processing of outputs from two SNSPDs fabricated on the same chip as the logic gate. Moreover, we show that nTrons can reach voltage levels compatible with lithium-niobate modulator control (1.15\,V) in a preliminary capacitive-load test at 100\,kHz. Although speed optimization remains necessary, the result potentially extends previous lower-voltage SNSPD–modulator interfaces \cite{deCea:20} toward fully on-chip feedforward operation.

The circuit schematic of the reconfigurable gate coupled to two independent detectors (SNSPD$_1$ and SNSPD$_2$) is shown in Fig.~\ref{fig:1}\textcolor{blue}{(a)}. If photons are detected, the SNSPDs drive the side nTrons (nT$_1$ and nT$_2$), which serve as amplifiers and buffers, into the resistive state. When these nTrons switch, fractions of their bias current (\(i_1\) and \(i_2\))  are redirected toward the gate of the central nTron (nT$_3$), where they add to give \(i_3 = i_1 + i_2\). If \(i_3\) exceeds the gate critical current $I_\text{th}$ of nT$_3$, an output pulse is generated. 
Shunt resistors ($R_\text{sh}$) ensure nTrons always reset to the superconducting states without latching, and kinetic inductors ($L_1$) reduce cross-talk between nT$_1$ and nT$_2$ when they fire. Resistors \(R_1\) provide a resistive path for current relaxation of $i_1$ and $i_2$, avoiding the buildup of circulating currents under repeated logic operations.
The circuit can be reconfigured to perform XOR, AND, and OR logic operations by tuning the bias currents of nT$_1$ and nT$_2$ ($I_\text{b1}$ and $I_\text{b2}$), while the current of nT$_3$ ($I_\text{b3}$) is kept close to its channel switching threshold.

We determined the target circuit parameters of the device in LTspice simulations, using the electro-thermal models of nTrons\cite{castellani2020design} and SNSPDs\cite{BerggrenSPICE}.
Fig.~\ref{fig:1}\textcolor{blue}{(d)} shows the simulated temporal evolution of \(i_1\), \(i_2\), and \(i_3\), with the corresponding output voltage \(v_{\mathrm{OUT}}\), for the XOR and AND configurations (OR operation shown in Supplementary Section~1). 

In the AND configuration, the two side nTrons are biased with the same current $I_\mathrm{b1} = I_\mathrm{b2} = I_\mathrm{L}$. When a single input pulse arrives, the current diverted to $i_3$ is insufficient to switch nT$_3$. When both inputs fire simultaneously, corresponding to bits $(1,1)$, the two contributions add constructively, giving $i_3 = i_1 + i_2 > I_\text{th}$ and producing an output pulse. In the XOR configuration, nT$_1$ and nT$_2$ are biased symmetrically near their switching currents but with opposite polarity ($I_\text{H}$ and $-I_\text{H}$) with $\lvert I_\mathrm{H}\rvert > \lvert I_\mathrm{L} \rvert$. When either nT$_1$ or nT$_2$ fires, corresponding to input combinations $(1,0)$ or $(0,1)$, an output pulse on nT$_3$ is generated. When simultaneous photons arrive, the opposite bias contributions cancel $\bigl(i_3 = i_1 + i_2 \approx 0\bigr)$, preventing nT$_3$ from switching. In the OR configuration $\bigl(I_\text{b1} = I_\text{b2} = I_\text{H}\bigr)$, inputs $(1,1)$ add constructively and also trigger switching.

Fig.~\ref{fig:1}\textcolor{blue}{(c)} illustrates an envisioned application of this logic gate to photonic feedforward processing with a beam-splitter–based interferometric setup. 
Thanks to quantum interference, two input photons exit from a beam splitter (directional coupler), either bunched into the same output port or separated into different ports. Detection events at D$_1$ and D$_2$, implemented with waveguide-integrated SNSPDs, are processed by the reconfigurable logic gate.
Assuming that no photons are lost, that no unwanted photons are introduced in this process, and that intrinsic dark counts are negligible, the AND operation selectively identifies coincidence events, while the XOR operation selects bunching events. More sophisticated setups, not implemented here, would combine photon-number resolving detectors to avoid misclassifications due to photon losses (e.g., zero photons on $D_1$ and one on $D_2$ classified as zero photons on $D_1$ and two on $D_2$).
The circuit output controls an optical switch on a separate waveguide where a third photon propagates with an engineered optical delay to ensure synchronization between the circuit signal and photon arrival.

Fig.~\ref{fig:1}\textcolor{blue}{(b)} shows the scanning electron micrograph of the fabricated logic gate with integrated resistors. 
We patterned the resistors ($R_1$ and $R_\text{sh}$) by direct-write photolithography and lift-off of 15\,nm gold with a 5\,nm titanium adhesion layer deposited by electron-beam evaporation on a 300\,nm SiO$_2$/Si substrate \cite{castellani2020design}. We subsequently sputtered a 10\,nm thick NbN film \cite{Castellani2024Counter} in direct contact with the resistors. On this layer, we patterned the superconducting components, including nTrons, inductors, detectors, resistor pads, and interconnects, using electron-beam lithography and CF$_4$ reactive ion etching.
We implemented the inductors as meandered 600\,nm wide nanowires, yielding inductances of $L_d \approx 8$\,nH and $L_1 \approx 10$\,nH. The nTrons featured a 43\,nm wide choke and a 334\,nm wide channel, while the SNSPDs consisted of 100\,nm wide meandered wires with a 33\% fill factor. We patterned all the devices on a single $1 \times 1$\,cm$^2$ chip, with individual circuit footprints of approximately $80 \times 200$\,µm$^2$ and SNSPD footprint of $11 \times 13$\,µm$^2$. 
Close-up views of the circuit components and detectors are shown in Supplementary Sections~2 and 5. Devices from a single fabrication run operated successfully on the first cooldown.

We experimentally verified that the gate correctly implements reconfigurable logic operations on sequences of 5\,ns wide input pulses generated by an arbitrary waveform generator (AWG) in periodic bursts, with an inter-pulse time of 200\,ns and a burst period of 1\,µs. Fig.~\ref{fig:ops}\textcolor{blue}{(a)} shows a simplified schematic of the setup (more details in Supplementary Section~3), and Fig.~\ref{fig:ops}\textcolor{blue}{(b)} shows the time-domain output traces for the same circuit configured as an AND, XOR, and OR gate. The input traces correspond to the AND measurement, while the XOR and OR outputs are shown together to illustrate correct logic and reconfigurable operation.

To characterize timing sensitivity, we investigated how the inter-arrival time between two AWG-generated input pulses, designed to emulate SNSPD output pulses, affects AND and XOR operations for the $(1,1)$ input state. This timing dependence is particularly important for coincidence detection with SNSPDs, where, depending on the quantum protocol, the encoding scheme of the quantum information, and optical delays of the setup, either short or long inter-arrival times may be expected for correlated pairs. We evaluated the operation success probability over 1000 trials as a function of the inter-arrival time, as shown in Fig.~\ref{fig:ops}\textcolor{blue}{(c)}.
To control the relative timing, the frequency of the input pulse sequences was slightly detuned between the two channels to allow for measurement of operation with sub-nanosecond phase increments.

\begin{figure}[ht!]
    \centering
    \includegraphics[width=0.9\linewidth]{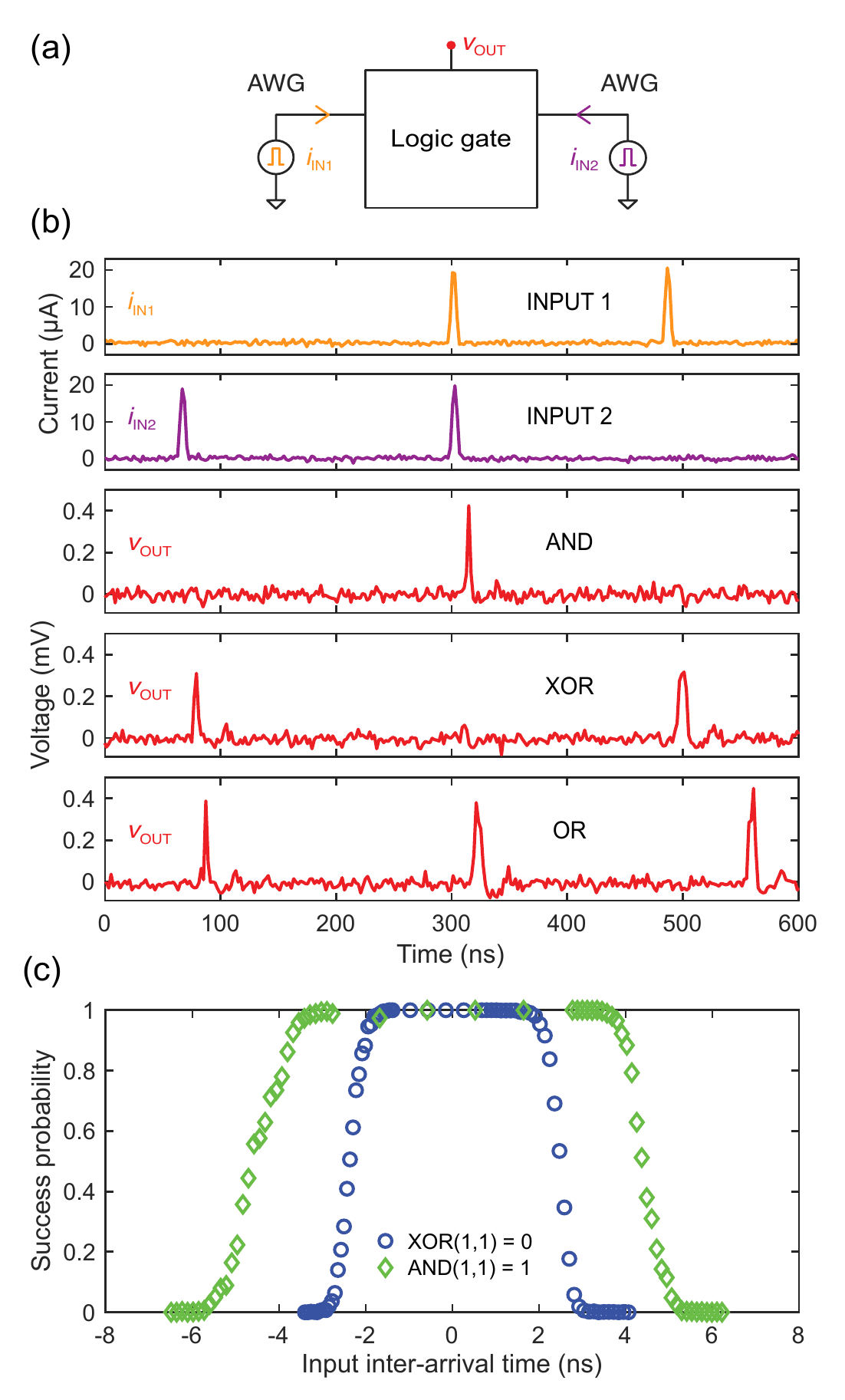} 
    \caption{Experimental reconfigurable logic operation and pulse timing dependence at 4.2\,K.
    (a) Simplified schematic of the logic gate tested with two input trains of $5\,\mathrm{ns}$-wide pulses generated by an arbitrary waveform generator (AWG).
    (b) Time-domain output traces for the logic gate configured for AND ($I_\text{b1,2}=60$\,µA), XOR ($I_\text{b1}=86$\,µA and $I_\text{b2}=-96$\,µA), and OR ($I_\text{b1,2}=100$\,µA) operations with $I_\text{b3}=99$\,µA. The two input traces correspond to the signals used during the AND-gate measurement. For the XOR and OR operations, the input pulses (not shown) had slightly different relative phases, leading to different timing in output pulses between the three gate operations. The output responses for AND, XOR, and OR are shown in the three output panels, respectively. The voltage amplitudes are divided by the nominal gain of the amplifiers at 1\,GHz.
    (c) Success probability of AND and XOR operations on input $(1,1)$ as a function of the inter-arrival time between the two input pulses.}
    \label{fig:ops}
\end{figure}

Both AND and XOR gates exhibit sharp decreases in success probability when inter-arrival times exceed a few nanoseconds. The AND gate maintains a probability higher than 90\% within a relative delay window of $-3.7\,\text{ns}$ to $+4\,\text{ns}$ (span: $7.7\,\text{ns}$), while the XOR gate tolerates $\pm 2.05\,\text{ns}$ (span: $4.1\,\text{ns}$). The probability drops to zero at approximately $\pm6$\,ns for the AND gate and $\pm3$\,ns for the XOR gate. These windows can, in principle, be adequate for HOM-style measurements with pump repetition rates lower than 167\,MHz ($\sim 6$\,ns$^{-1}$) for the AND gate, and 333\,MHz ($\sim 3$\,ns$^{-1}$) for the XOR gate. Higher values could generate false coincidence-detection events between photons detected in consecutive laser pulses.

The time window values depend on the pulse shapes and circuit dynamics in the internal branches of the circuit ($i_1$, $i_2$, and $i_3$); in particular, longer $L/R$ reset time constants of nT$_1$ and nT$_2$ increase the time windows by prolonging $i_1$ and $i_2$. Therefore, tuning the resistances and inductances allows the time windows to be controlled according to the application.
The observed discrepancy between XOR and AND time windows is consistent with our simulations and can be attributed to the asymmetric shape of the nTron output pulses in $i_1$ and $i_2$. In the AND configuration, the fast rising edge of the later pulse can overlap with the long falling edge of the earlier pulse, allowing $i_3$ to exceed $I_\text{th}$ for extended inter-arrival delays. By contrast, in the XOR configuration, the later inverted pulse must arrive very close to the rising edge of the first pulse to cancel it out before $I_\text{th}$ is reached. Therefore, the XOR gate is more sensitive to timing mismatch, resulting in a narrower operation window.

\begin{figure*}[t]
    \centering
    \includegraphics[width=1\linewidth]{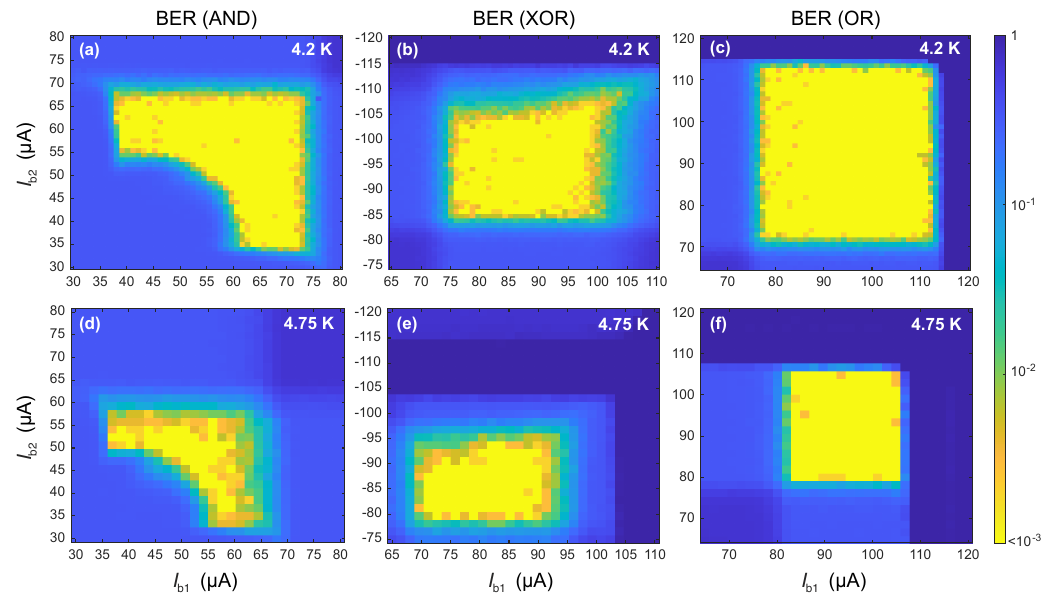}
    \caption{Bit error rate (BER) as a function of bias currents $I_\text{b1}$ and $I_\text{b2}$ for (a, d) AND, (b, e) XOR, and (c, f) OR configurations. Panels (a-c) were obtained at $4.2\,\mathrm{K}$ with $I_\text{b3} = 99$\,µA and panels (d-f) at $4.75\,\mathrm{K}$ with $I_\text{b3} = 93$\,µA. The input peak current was 20.5\,µA, and the inter-arrival time was set to zero for all the data points. The input signals were bursts of three different pair configurations with an inter-pulse time of 200\,ns and a burst period of 1\,µs. Bright yellow regions indicate completely correct results, and dark blue regions indicate completely wrong results over $10^{3}$ logic trials.}
    \label{fig:BER}
\end{figure*}

Following the timing analysis, we measured the bit-error rate (BER) as a function of the two side-nTron bias currents ($I_\text{b1}$, $I_\text{b2}$) at 4.2\,K for all logic configurations, on the same input pattern of Fig.~\ref{fig:ops}\textcolor{blue}{(b)}.
Fig.~\ref{fig:BER}\textcolor{blue}{(a-c)} shows the resulting margins, where bright yellow regions correspond to a BER below $10^{-3}$, computed over 1000 operations per bias point. 
A description of the analysis procedure is provided in Supplementary Section~4.

The OR gate exhibits an approximately square window of correct operation in Fig.~\ref{fig:BER}\textcolor{blue}{(c)}, with slightly different bias ranges: $I_\text{b1}=(94.5\pm17.5)\,\text{µA}$ ($\pm18.5\%$) and $I_\text{b2}=(92\pm20)\,\text{µA}$ ($\pm21.7\%$). At the lower and left boundaries of this window, the bias currents are insufficient to trigger switching of nT$_3$. Beyond the upper and right boundaries, excessive bias currents cause the nTrons to latch. 
In contrast, the AND gate exhibits an L-shaped operating region in Fig.~\ref{fig:BER}\textcolor{blue}{(a)}. Along the left edge ($I_\text{b2}=55$--67\,\textmu A) and the lower edge ($I_\text{b1}=61$--72\,\textmu A) of the yellow region, a sufficiently high bias on one input can compensate for a lower bias on the other. In the bottom left region, the combined contribution of the two bias currents is insufficient to trigger switching. The largest square area inside the L-shaped region is defined by $I_\text{b1}=(63.5\pm8.5)\,\text{µA}$ ($\pm13.4\%$) and $I_\text{b2}=(58.5\pm8.5)\,\text{µA}$ ($\pm14.5\%$). 
Finally, in the XOR configuration (Fig.~\ref{fig:BER}\textcolor{blue}{(b)}), the operating window is rectangular with a more pronounced asymmetry than in the OR mode: $I_\text{b1}=(87\pm11)\,\text{µA}$ ($\pm12.6\%$) and $I_\text{b2}=(-94.5\pm9.5)\,\text{µA}$ ($\pm10.1\%$). 
For comparison, other cryotron-based circuits have exhibited similar or higher levels of robustness with bias margins on the order of 12--40\% \cite{Buzzi2024Nanocryotron, Foster2023ShiftRegister, Castellani2024Counter}.

The window asymmetry in the XOR configuration likely arises from the application of positive input currents to both side nTrons while one of them is negatively biased. 
Moreover, the top and right boundaries are not straight as in the OR mode. Correct operation extends into the top-right corner, with a noticeable curvature of the edges. 
Along these edges, there is an intermediate light-blue region before the transition to the dark-blue area, where the nTrons latch. In this transition region, for part of the (1,1) input operation, the output current from one of the two side nTrons is insufficient to fully cancel the contribution of the other. 
Moreover, in proximity to the top and right edges of the XOR and OR windows, the circuit occasionally produces two consecutive output pulses instead of one. Such behavior is not considered an error in these maps.
For all gate configurations, device-level asymmetries between the two branches, such as variations in choke width, kinetic inductance, or resistance, may contribute to the margin imbalances. 

To assess the robustness of the bias margins against temperature variations, we extracted the margins at 4.75\,K (0.55\,K variation). As shown in Fig.~\ref{fig:BER}\textcolor{blue}{(d-f)}, the operating regions shift toward lower absolute bias currents and become narrower, with the AND gate exhibiting the strongest margin contraction of 40\%. XOR and OR margins contract by 10\% and 35\%, respectively.

With periodic input pulses at frequencies of 15\,MHz and 25\,MHz, the margins for BERs lower than $\sim10^{-3}$ become narrower, and asymmetries in the window shapes emerge, as shown in Supplementary Section~4. 
The speed limitation likely arises from the input signal period being comparable to the $L/R$ time constants of the circuit inductors, while asymmetries are probably introduced by parameter variations in the fabricated elements.
Optimized readout setup and $L/R$ values should enable frequencies above $25\,\mathrm{MHz}$, 
supported by prior nanowire-based circuit results: shift-register operation at $200\,\mathrm{MHz}$ 
\cite{foster_time-tagging_2025} and nTron switching up to $615\,\mathrm{MHz}$ 
\cite{Zheng2019SciRep_nTron615MHz}.
Ultimately, the intrinsic speed limit of this technology is set by the thermal reset time of the hotspot, which depends on the thermal interface to the substrate and could therefore be improved through substrate engineering \cite{incalza_fast-recovery_2026}.

\begin{figure}[ht!]
    \centering
    \includegraphics[width=0.9\linewidth]{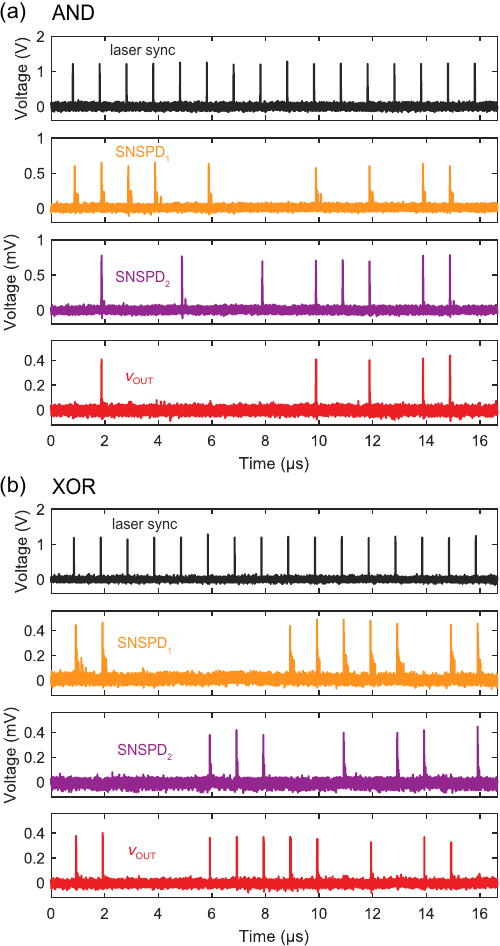} 
    \caption{Time-domain experimental traces of the reconfigurable logic gate operated in two configurations: (a) AND mode ($I_\text{bs1,2} = 21$\,µA, $I_\text{bg1,2} = 14$\,µA, $I_\text{b1} = 83$\,µA, $I_\text{b2} = 87$\,µA, $I_\text{b3} = 110$\,µA) and (b) XOR mode ($I_\text{bs1} = 21$\,µA, $I_\text{bs2} = -21$\,µA, $I_\text{bg1} = 10$\,µA, $I_\text{bg2} = -4$\,µA, $I_\text{b1} = 94$\,µA, $I_\text{b2} = -106$\,µA, $I_\text{b3} = 105$\,µA). In each case, the panels show (from top to bottom) the 1\,MHz sync signal of the pulsed laser, the amplified output pulses of SNSPD$_1$ and SNSPD$_2$, and the amplified circuit output voltage. For the XOR operation, the output pulses of SNSPD$_2$ have negative amplitude (which yielded a lower BER), but the signal is inverted for clarity. The detector and output voltage amplitudes are divided by the amplifiers' nominal gain at 1\,GHz.}
    \label{fig:snsdp_integration}
\end{figure}

Although this experiment uses AWG-generated input pulses, the intended application of this gate is to interface with SNSPDs. The lowest verified BER of $\sim10^{-3}$ in Fig.~\ref{fig:BER} would likely be negligible at the system level, since SNSPDs typically provide 90--99\% detection efficiency. If required, lower BER values are achievable with this logic gate: with further optimization of the bias point, we reduced the BER below $10^{-5}$ at 4.2\,K, but we did not extract a full BER map.

To move toward the envisioned use of the circuit in on-chip Bell-state measurements and feedforward schemes, we evaluated its logic operation on output pulses from SNSPDs fabricated on the same chip as the circuit. 
We connected two detectors through wire-bonds and surface-mount resistors, using the same configuration as in Fig.~\ref{fig:1}\textcolor{blue}{(a)} (with a different $R_2=2$\,$\Omega$). However, we added two bias lines to the chokes of the side nTrons ($I_\text{bg1}$ and $I_\text{bg2}$) to tune their effective switching thresholds and thus maximize switching probability. More details about the setup are shown in Supplementary Section~3.
Due to the thickness of the film (10\,nm), the SNSPDs did not operate with saturated internal detection efficiency at 1550\,nm (Supplementary Section~5 shows characterization results of the detectors).

The detectors were flood illuminated at 1550\,nm using a pulsed laser diode (PicoQuant LDH-P-C-1550) operating at a 1\,MHz repetition rate.
To increase the probability of coincidence events, no optical attenuation was applied, resulting in operation in a multi-photon detection regime. SNSPD operation in the single-photon regime (discussed in Supplementary Section~5) would not alter the circuit dynamics.
Detector and circuit outputs were simultaneously recorded, allowing direct implementation and validation of logic operations on SNSPD-generated signals.

Fig.~\ref{fig:snsdp_integration} shows the time-domain experimental traces of the integration scheme, which reproduces the expected AND and XOR behaviors for 16 consecutive laser pulses. At the optical power used, each laser pulse randomly generates one of the four input bit-pair combinations, with comparable probability.
We extracted the BER of the detector-driven circuit using $10^4$ laser pulses, excluding operations corresponding to $(0,0)$ detection events (approximately $20\%$ of the total operations). We obtained BERs of $2.43 \times 10^{-2}$ for the AND mode and $3.19 \times 10^{-2}$ for the XOR mode under the optimal bias conditions (values in the caption of Fig. \ref{fig:snsdp_integration}).
The optimal bias points for both operations lie outside or at the boundaries of the margins obtained for the AWG-driven gate in Fig.~\ref{fig:BER}. This discrepancy arises from the shorter input pulses and the lower output impedance of the shunted detector compared to the 10 k$\Omega$ input bias resistor of the previous experiment. 

The BER values are higher than those obtained with the AWG-driven gate, with increased uncertainty in the output-pulse arrival time required to maintain these BER values. For the AND operations, the minimum error rate is achieved by accepting as valid any output pulse arriving with a delay lower than 19\,ns relative to the SNSPD pulses. 
For XOR operations, this delay increases to 26\,ns. 
Shorter delays correspond to higher bit-error rates (see Supplementary Section~6 for more details).
We note the presence of afterpulses in the SNSPD output traces, which contribute to the increased timing uncertainty and higher BER. Afterpulsing is predominant on SNSPD$_1$, but it is also visible on SNSPD$_2$ (close-up views of the circuit pulses are shown in Supplementary Section~6). This asymmetry is consistent with differences in the wire-bonding interconnects between SNSPDs and nTrons or possible device-to-device parameter mismatch. 

We attribute afterpulsing, timing uncertainty, and the BER increase to a sub-optimal electrical coupling between detectors and nTrons. 
In particular, the 50\,$\Omega$ coaxial line used to read out the detector's signal diverts part of the SNSPD output current that would otherwise be delivered to the nTron gate, introduces impedance-mismatch reflections, and might contribute to unwanted relaxation oscillations of both the detector and the nTron choke.

In a practical use case of the system, the 50\,$\Omega$ readout line used to acquire SNSPD pulses would be disconnected. Moreover, the inductances and resistances between the detector and the nTron, which govern the relaxation-oscillation dynamics, could be optimized in a fully integrated design (without interconnections through wire bonds).

The BER values reported here are still comparable to the typical system detection error rates of well-optimized SNSPDs. In addition, they do not represent the intrinsic limits of the technology: substantially lower BERs have been achieved in the past when detectors were optimally coupled to superconducting circuits \cite{Castellani2024Counter}.

We estimated the circuit's energy consumption by reproducing the experimental operating conditions of Fig.~\ref{fig:snsdp_integration} in LTspice simulations.
For both logic configurations and all input combinations, the energy per operation ranged from $\sim 110$\,aJ ($\text{AND}(1,0)=0$) to $\sim 315$\,aJ ($\text{AND}(1,1)=1$).
When accounting for the full system, including the SNSPDs, the energy per operation was between $\sim\!165$\,aJ and $\sim\!450$\,aJ (see Supplementary Section~7 for more details).
These values are comparable to switching energies previously reported for nTron logic \cite{Buzzi2024Nanocryotron} and ripple counters \cite{Castellani2024Counter}. 
Our values exclude the energy dissipated by off-chip 10\,k$\Omega$ bias resistors. This overhead is not fundamental and can be minimized through lower-resistance and inductive bias distribution strategies in large-scale circuits~\cite{kirichenko2011zerostatic}.

Future work will focus on improving the performance of the circuit driven by detectors through optimized fabrication, enhanced electrical coupling, and reduced afterpulsing, as well as on improving detector efficiency at 1550\,nm via thinner superconducting films or refined designs \cite{incalza_fast-recovery_2026}. Alternatively, separating detection and computation into distinct layers could enable independent optimization at the cost of added fabrication complexity.

Realizing fully integrated feedforward control requires cryogenic logic that can drive electro-optic modulators to their control-voltage levels. To assess this capability, we fabricated a wide-channel nTron (1.5\,\textmu m channel, 200\,nm choke, 10\,nm thickness) intended as an output-stage amplifier for the logic gate. At 4.2\,K, the device charged a 1\,pF capacitive load up to 1.15\,V, using 5\,ns wide, 68\,\textmu A input pulses at 100\,kHz; the load approximates the low-frequency impedance of a gate-controlled electro-optic modulator.
These output levels and speeds are comparable to results reported in the literature \cite{Zhao2017nTronComparator,paul_photolithography-compatible_2025}. Lower bias enabled higher repetition rates at reduced output voltage by reducing Joule heating. These results suggest that nTrons can provide sufficient drive for cryogenic electro-optic or CMOS-compatible elements, supporting their use in quantum feedforward. 
However, device geometry, substrate material, and deposition processes should be 
optimized to enhance thermal recovery~\cite{incalza_fast-recovery_2026} and speed, while 
maintaining compatibility with photonic substrates.
Further details and discussions are given in Supplementary Section~8.

In conclusion, the demonstrated bias-configurable nanocryotron logic gate provides a compact and energy-efficient building block for on-chip superconducting logic compatible with SNSPD technology. By tuning bias currents, a single circuit can implement AND, OR, and XOR functions with sub-$10^{-3}$ error rates when driven by AWG-generated signals, offering flexible operation within the same layout. This reconfigurability simplifies cryogenic logic design and enables direct integration with superconducting detector arrays. The results shown for XOR and AND operations on SNSPD outputs demonstrate the potential of co-integrating nanocryotron electronics and detectors. However, additional work will be required to improve the coupling between the two systems, and thereby reduce the output timing uncertainty and bit-error rates.
Finally, the nTrons output voltages, compatible with electro-optic modulators (shown in the supplementary material), suggest the feasibility of driving switching or frequency-shift operations in the optical domain.
Future work will extend these concepts toward waveguide-integrated SNSPDs coupled to on-chip nTron circuits that can directly interface with modulators, paving the way for scalable quantum feedforward control. More broadly, demonstrations of multi-gate logic functions will benefit low-power signal processing in extreme environments. 

Before the submission of this paper, we became aware of a recent, similar implementation of an AND gate for coincidence detection based on five nTrons~\cite{liu_tunable_2026}. While our work emphasizes logic reconfigurability, that work focuses on tuning the inter-arrival time window for coincidence detection.

\vspace{-0.24cm}
\begin{acknowledgments}

The initial part of this work was supported by the National Science Foundation under Grant No. OMA-2137723 and the Center for Quantum Networks (CQN) Grant No. EEC-1941583. The second stage of this work was sponsored by the U.S. Department of Energy (DOE), Office of Science, Office of Basic Energy Sciences, under Award No. DE-AC02-07CH11359. Additional support was provided by the DOE Office of Science, Offices of High Energy Physics and of Basic Energy Sciences, as part of the Co-design and Heterogeneous Integration in Microelectronics for Extreme Environments (CHIME) Microelectronics Science Research Center (MSRC), under Contract No. LAB 24-3320. Alejandro Simon acknowledges support from the NSF GRFP. We thank Ari Willner and Davide Mondin for their careful review of the manuscript and for their helpful comments and suggestions. We thank Adam McCaughan for contributing to the early-stage ideation of the XOR gate, Neil Sinclair for discussions on photonic feedforward operations, and Whitney R. Armstrong and Davide Braga for discussions on the pulse inter-arrival time.
\end{acknowledgments}

\vspace{-0.24cm}
\section*{Data Availability Statement}

The data that support the findings of this study are available
from the corresponding author upon reasonable request.

\vspace{-0.24cm}
\section*{References}
\vspace{-0.24cm}

\bibliography{gate_bib}

\end{document}

\begin{center}
    \vspace*{2cm}
    {\LARGE \textbf{Supplementary Material}}\\[0.5cm]
    {\Large {Reconfigurable Superconducting Logic for On-Chip Photon Coincidence Detection}}\\[1cm]
    {\large Gabriel Le Guay, Matteo Castellani, Reed Foster, Francesca Incalza, Alejandro Simon, Owen Medeiros, Phillip D. Keathley, and Karl K. Berggren}\\[0.5cm]
    Research Laboratory of Electronics, Massachusetts Institute of Technology,\\
    Cambridge, Massachusetts 02139, United States\\[2cm]

\end{center}

\newpage
\supplementarysection
% ===== TABLE OF CONTENTS =====
\tableofcontents
\newpage

\section{Simulation of the OR operation}

\begin{figure}[H]
    \centering
    \includegraphics[width=0.55\linewidth]{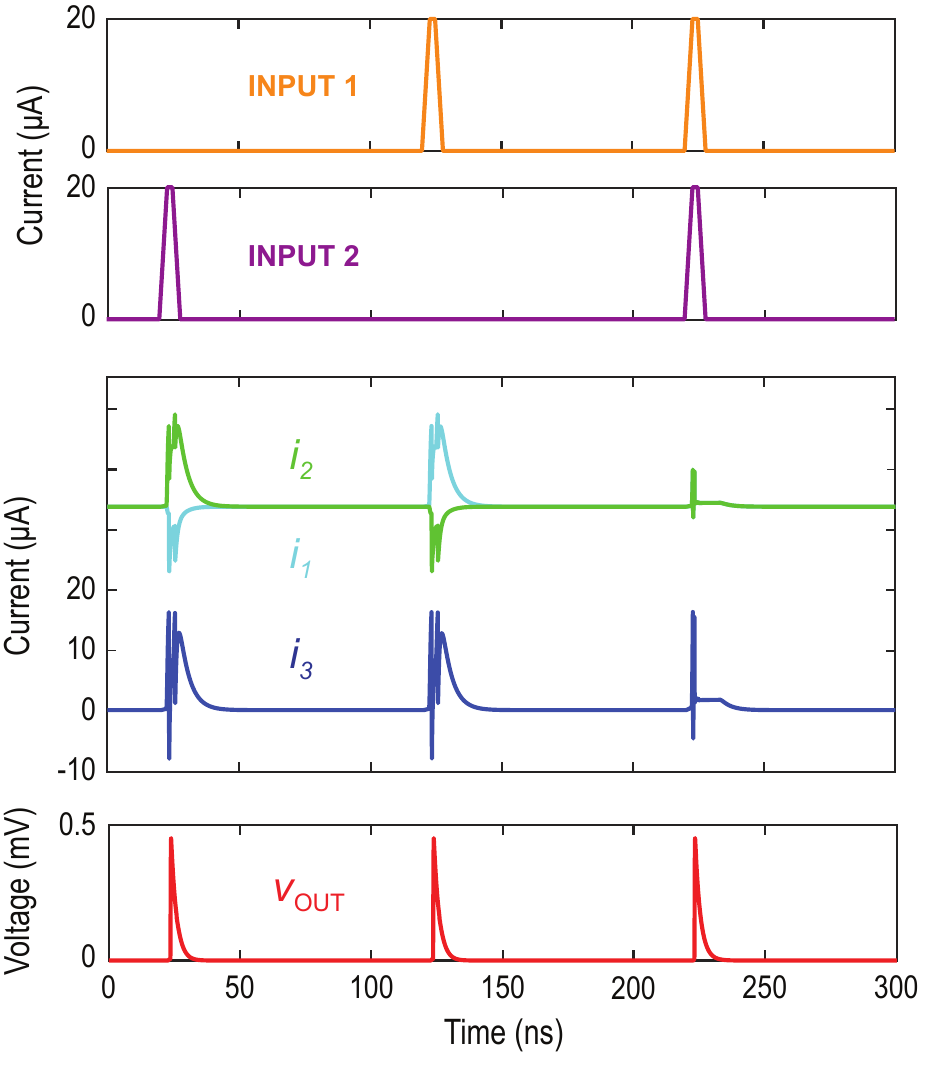}
    \caption{Time-domain simulation of the gate dynamics for the OR configuration. From top to bottom, the panels show the input pulses applied to the gates of nT$_1$ and nT$_2$ (INPUT 1 and INPUT 2), the corresponding internal currents \(i_1\) and \(i_2\), their sum \(i_3\), and the output voltage \(V_{\mathrm{out}}\). For visual clarity and compactness, the current traces \(i_1\) and \(i_2\) are vertically shifted along the current axis.}
    \label{fig:OR_SIM}
\end{figure}

Fig.~\ref{fig:OR_SIM} presents the simulated temporal response of the logic gate operating in the OR mode, when driven by ideal current sources. The input pulses applied to the gates of nT$_1$ and nT$_2$ trigger corresponding pulses in \(i_1\) and \(i_2\), which combine additively to drive the central branch current \(i_3\). In this configuration, the gate switches whenever at least one of the two inputs is active, producing a voltage output \(v_\text{OUT}\) for every logical event.

The double-spike feature in $i_3$ for inputs $(0,1)$ and $(1,0)$ arises because the gate of nT$_3$ switches twice, even though its channel switches only once. This occurs because when only one side nTron fires and the gate hotspot resets quickly enough to trigger a second switching event. The subsequent slow decay in $i_1$, $i_2$, and $i_3$ results from the release of energy stored in the inductors.
While this double-spike behavior does not affect the output, it may cause instabilities at higher operating frequencies. The effect is more pronounced in the XOR and OR configurations, where the side nTron bias currents are higher than in the AND case. It can be suppressed by lowering the bias currents toward the bottom of the operating margins or by increasing the $L/R$ time constant of the two circuit branches.

\section{Details of the fabricated circuit components}

\begin{figure}[h!]
    \centering
    \includegraphics[width=0.7\textwidth]{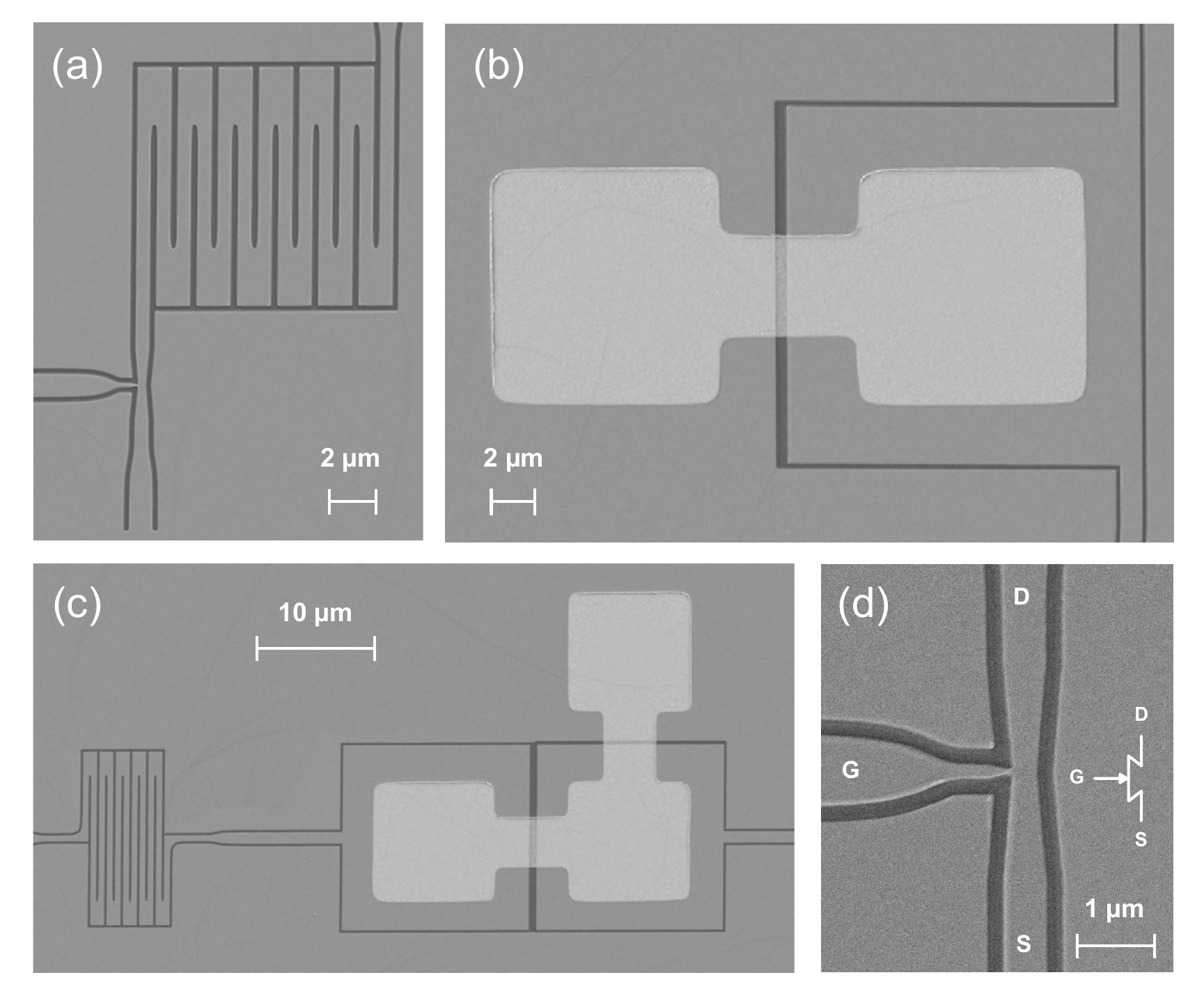}
    \caption{
        \textbf{(a)} Scanning electron micrograph of the nTron with its drain kinetic inductor $L_\text{d}$ implemented as a meandered nanowire (8\,nH). 
        (b) SEM of the Ti/Au shunt resistor ($R_\text{sh} = 5\,\Omega$) and its contact pad. 
        (c) Series combination of a kinetic inductor ($L_1 = 10\,\mathrm{nH}$) and resistor ($R_1 = 10\,\Omega$) connected in parallel to a shunt resistor ($R_\mathrm{sh} = 5\,\Omega$). 
        (d) Detailed SEM of the nanocryotron (nTron) with its circuit symbol; the 43\,nm wide choke connects the gate (G) to the 334\,nm wide channel between the drain (D) and source (S).
    }
    \label{fig:SEM}
\end{figure}

Fig.~\ref{fig:SEM} shows scanning electron micrographs (SEMs) of the main components constituting the bias-configurable superconducting logic gate. The nanocryotron (nTron) is shown together with its drain kinetic inductor $L_\text{d}$, implemented as a meandered nanowire providing an inductance of 8\,nH. Fig.~\ref{fig:SEM}(b) displays the patterned Ti/Au load resistor ($R_L = 5\,\Omega$) with its corresponding contact pad, large enough to provide a negligible contact resistance between NbN and gold. In Fig.~\ref{fig:SEM}(c), the series combination of a kinetic inductor ($L_1 = 10\,\mathrm{nH}$) and a resistor ($R_1 = 10\,\Omega$) is shown; this branch is connected in parallel to a shunt resistor ($R_\text{sh} = 5\,\Omega$) to stabilize the device operation and control the bias current distribution. The wire width of the inductors (600\,nm) was chosen large enough to keep the meanders always underbiased and avoid undesired photon detection events when the chip is flood illuminated. The sheet kinetic inductance of the film is approximately 40\,pH/$\square$.
Fig.~\ref{fig:SEM}(d) shows a close-up view of the nTron together with its circuit symbol. The device features three terminals: the drain (D), the source (S), and the gate (G). It includes a 43\,nm wide choke that connects the gate to the 334\,nm wide channel.

 The critical temperature was not directly extracted for this chip; however, 10\,nm thick films deposited with the same process typically have values of $T_c \approx 8.5$\,K.
% ===== SECTION S1 =====
\section{Experimental setup and measurement procedure}

\begin{figure}[H]
    \centering
    \includegraphics[width=0.95\linewidth]{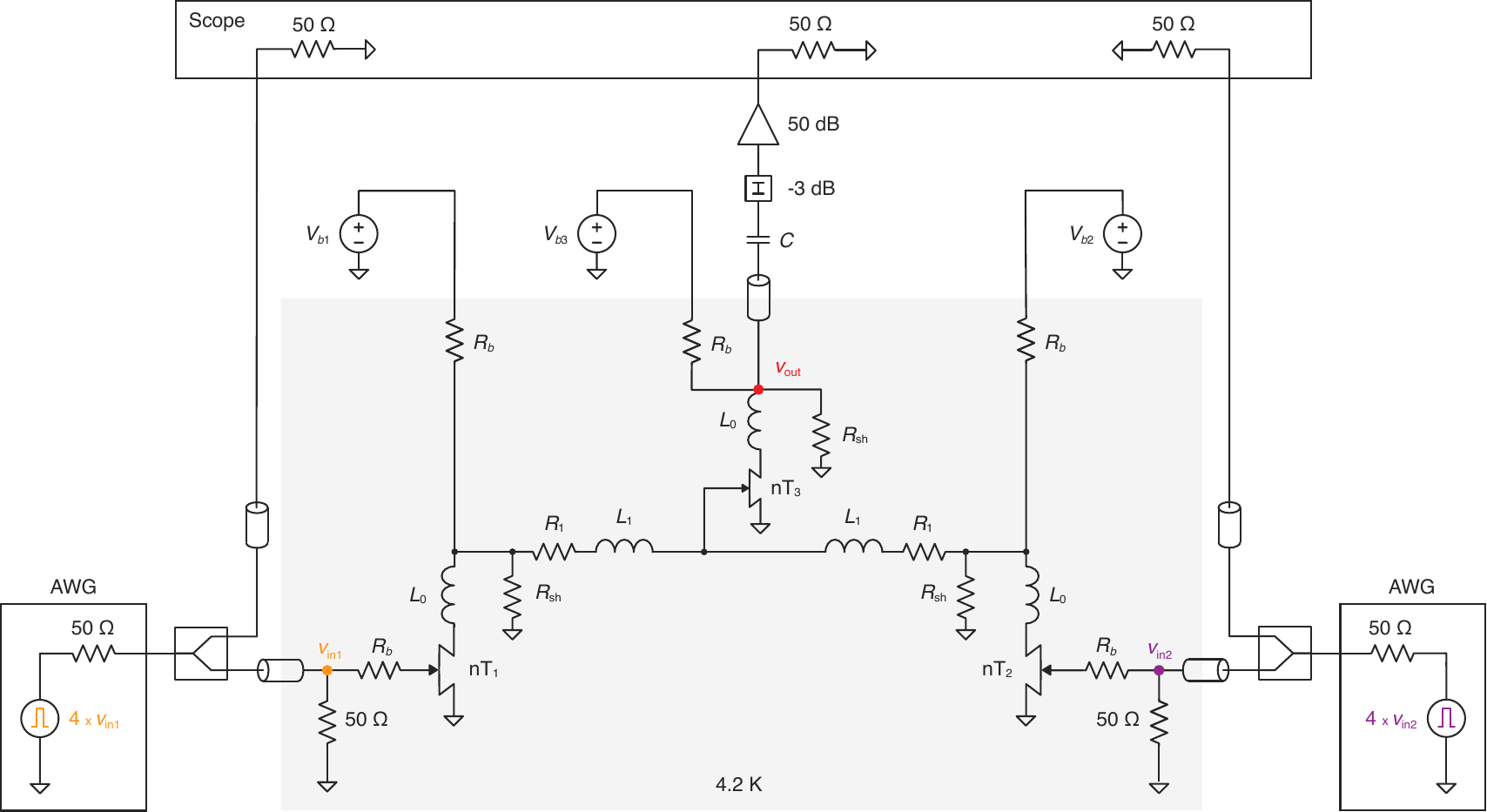} 
    \caption{Setup for the time-domain measurement of the reconfigurable gate. The input signals, generated by an arbitrary wave-form generator (source 1 and 2 of the AWG), are split and read by the oscilloscope as well as being sent to the device ($v_\text{in1}$ and $v_\text{in2}$). Coaxial cables connect the device from 4.2\,K to room-temperature electronics. A 50-$\Omega$ resistor in front of the input bias resistor ($R_\text{b} = 10\,\text{k}\Omega$) ensures matching between the PCB and the coaxial cable. The components enclosed in the gray box are on the chip or PCB at 4.2\,K. The nTrons are biased with room-temperature batteries ($V_\text{b1}$, $V_\text{b2}$, and $V_\text{b3}$) and low-temperature resistors ($R_\text{b} = 10\,\text{k}\Omega$). Each output signal passes through a 3\,dB attenuator, a bias tee with DC port terminated, and two amplifiers (providing a total gain of 47\,dB), before going to the oscilloscope. $C$ is the internal capacitance of the bias tee. The impedance of the scope is set to 50\,$\Omega$.}
    \label{fig:example_image}
\end{figure}

Fig.~\ref{fig:example_image} shows the measurement setup used for the characterization of the reconfigurable nanocryotron logic gate. The fabricated chip was mounted on a gold-plated printed circuit board (PCB) using GE varnish, with electrical contact to the device pads provided by aluminum wire bonding. Measurements were carried out at 4.2\,K in a liquid-helium dewar using a custom-built dip probe~[1]. Input current pulses with a duration of 5\,ns were generated by an arbitrary waveform generator (AWG) and delivered to the device through coaxial lines. At the PCB level, each bias line from low-noise voltage sources was connected to the circuit input through a 10\,k$\Omega$ series resistor, while a 50\,$\Omega$ resistor was placed in parallel at each input node to ensure impedance matching. The output signal was routed through a bias tee with the DC port terminated, followed by a fixed 3\,dB attenuator and two cascaded low-noise broadband amplifiers (25\,dB gain and 2.5\,GHz bandwidth each), and recorded using a real-time oscilloscope with a 6\,GHz bandwidth. Each logic configuration (AND, OR, XOR) was characterized by sweeping the bias currents ($I_\text{b1}$, $I_\text{b2}$) and evaluating the bit-error rate (BER) over $10^3$ to $10^5$ operations. Temperature-dependent measurements were additionally performed in the range from 4.2\,K to 4.8\,K.

\begin{figure}[H]
    \centering
    \includegraphics[width=0.95\linewidth]{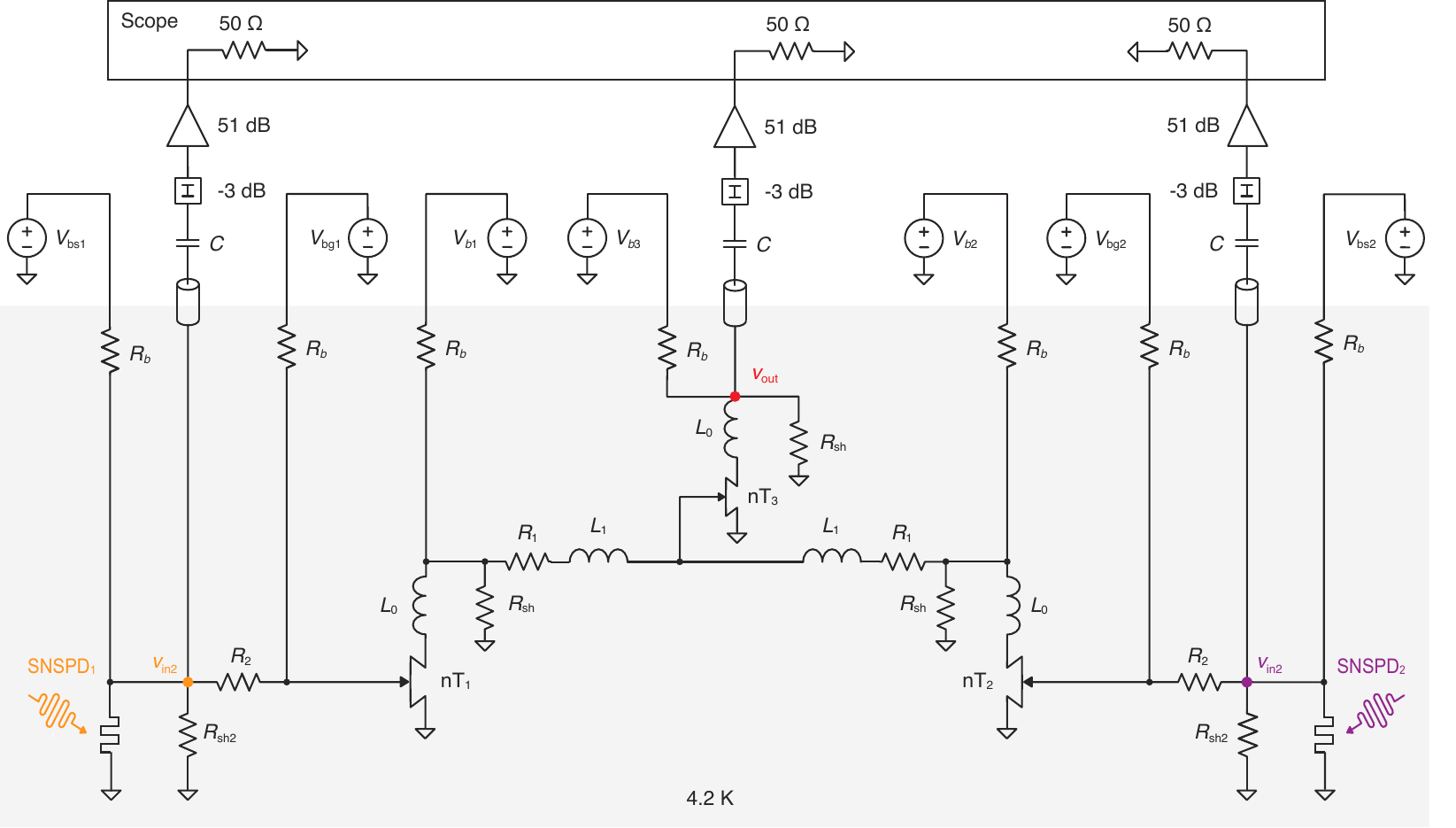} 
    \caption{Setup for the time-domain measurement of the reconfigurable gate coupled to SNSPDs. The SNSPDs are on the same chip as the circuit. Coaxial cables (50- impedance) connect the device from 4.2\,K to room-temperature electronics.  The components enclosed in the gray box are on the chip or PCB at 4.2\,K. The nTrons and SNSPDs are biased with room-temperature batteries ($V_\text{b1}$, $V_\text{b2}$, and $V_\text{b3}$, $V_\text{bs1}$, $V_\text{bg1}$, $V_\text{bs2}$, $V_\text{bg2}$,) and low-temperature resistors ($R_\text{b} = 10\,\text{k}\Omega$). Each output signal passes through a 3\,dB attenuator, a bias tee with DC port terminated, and two amplifiers (providing a total gain of 48\,dB), before going to the oscilloscope. $C$ is the internal capacitance of the bias tee. The impedance of the scope is set to 50\,$\Omega$.}
    \label{fig:setup_snspd}
\end{figure}

The measurement setup for the experiment with SNSPDs is illustrated in Fig.~\ref{fig:setup_snspd}.
Two identical detectors (SNSPD$_1$ and SNSPD$_2$) were respectively connected to the gates of nT$_1$ and nT$_2$, through 2\,$\Omega$ surface-mount resistors ($R_2$) and aluminum wire bonds. For each circuit input: a bias line for the detector was placed ($I_\text{bs1}$ and $I_\text{bs2}$); a 50\,$\Omega$ shunt resistor was added in parallel with the SNSPD to ensure self-reset when the nTron fires ($R_\text{sh2}$), and a bias line (current $I_\text{bg1}$ or $I_\text{bg2}$) on the nTron gate was added to tune its current threshold (This line was not used for simulations of Fig. 1 in the main text).
A 1550\,nm pulsed laser diode with a 1\,MHz repetition rate (PicoQuant LDH-P-C-1550), was used to flood illuminate both detectors with $\sim 0.5$\,mW of nominal output power. Optical attenuation was not applied in order to increase the probability of coincidence detection and facilitate data acquisition; as a result, the SNSPDs operated in a multi-photon detection regime.  
Detector and circuit outputs were simultaneously acquired using readout chains similar to that described above, except that the second amplifier had a 26\,dB gain and a 2\,GHz bandwidth.

\section{Methods for bit-error-rate (BER) estimation and results of BER frequency dependence}

The bit-error rate (BER) was extracted from time-resolved output traces recorded for each bias configuration. For every data file, multiple consecutive logic operations were acquired within a single trace, each operation corresponding to one of the three input states $(1,0)$, $(0,1)$, and $(1,1)$.
For each trace segment, the output signal $v_\text{OUT}$ was analyzed in the time domain. Three non-overlapping temporal windows of 1\,ns were defined, corresponding to the expected arrival times of the three logic operations. Within each window, a valid output event was registered when the signal amplitude exceeded a fixed threshold of 50\,\textmu V. Spurious pulses occurring outside the defined operation windows were detected using a slightly lower threshold to account for latching pulses.
An error was counted for a given operation if either (i) the expected output pulse was absent within its corresponding time window, or (ii) more than two spurious pulses were detected outside the valid windows. For each trace segment, errors were evaluated independently for the three operations, yielding up to three errors per segment.
The BER for a given bias configuration was then calculated as the total number of detected errors divided by the total number of evaluated logic operations. This procedure was repeated for all bias configurations, and the resulting BER values were mapped as a function of the two bias currents.

\subsubsection*{Frequency dependence}

\begin{figure}[H]
    \centering
    \includegraphics[width=1\linewidth]{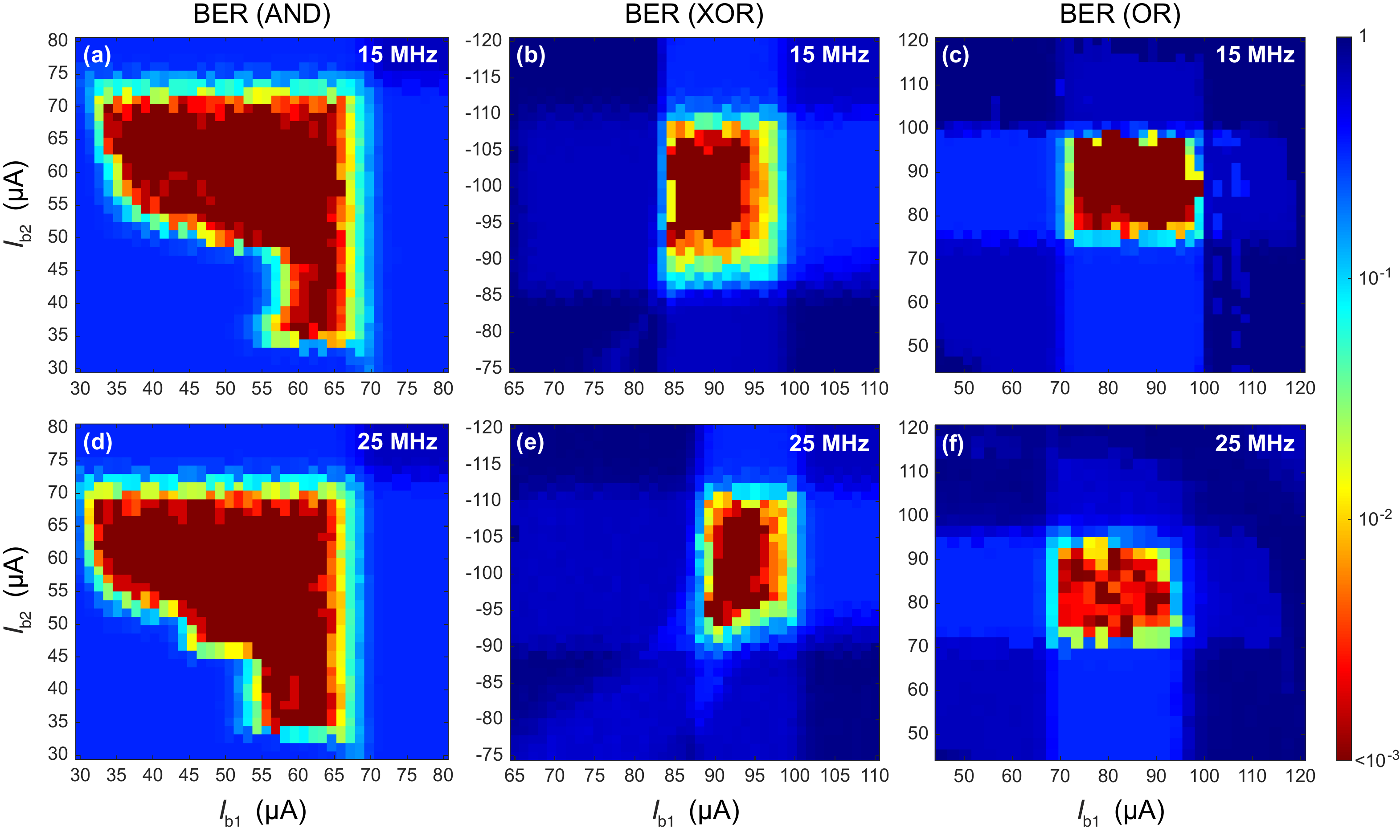}
    \caption{Bit--rate (BER) as a function of bias currents $I_\text{b1}$ and $I_\text{b2}$ for (a, d) AND, (b, e) XOR, and (c, f) OR configurations at $4.2\,\mathrm{K}$. Panels (a-c) were obtained with input pulses at 15\,MHz, and panels (d-f) at 25\,MHz, both with $I_\text{b3} = 99$\,µA. The input peak current was 20.5\,µA, and the inter-arrival time was set to zero for all the data points. Dark red regions indicate completely correct results and dark blue regions indicate completely wrong results over $10^{3}$ logic trials.}
    \label{fig:ber_freqs}
\end{figure}

Increasing the input frequency narrows the margins, as shown in Fig.~\ref{fig:ber_freqs}(a-c) at 15\,MHz and Fig.~\ref{fig:ber_freqs}(d-f) at 25\,MHz operation. The overall BER margins decreases with frequency, which might be caused by the electrical $L/R$ time constant of the circuit limits the full recharging of the branches within each cycle, as previously reported in similar nanowire circuits. Unlike the measurements above, these data were not acquired in burst mode, which changes the circuit dynamics and may also contribute to the reduced margins.

A notable feature is the emergence of bias asymmetry in the AND and XOR gates: at high frequency, the margins of \(I_\text{b1}\) become narrower than those of \(I_\text{b2}\). The origin of this effect remains unclear. In the XOR configuration, such asymmetry could arise from the opposite polarity of the bias and input pulse on the \(I_\text{b2}\) branch (negative bias with positive pulse), while on the \(I_\text{b1}\) side both are positive, potentially affecting the local switching dynamics and thermal reset time. For the AND gate, however, the operation is expected to be symmetric, making this behavior more puzzling. It may instead arise from extrinsic factors such as variations in wire bonding or fabrication imperfections, including unintended differences in choke width, kinetic inductance, or resistor values.

For the OR gate, the response remains largely symmetric, but the overall operation window shrinks, and the average BER inside the window for correct operation increases with frequency.

\newpage
\section{SNSPD characterization}

\begin{figure}[H]
    \centering
    \includegraphics[width=0.95\linewidth]{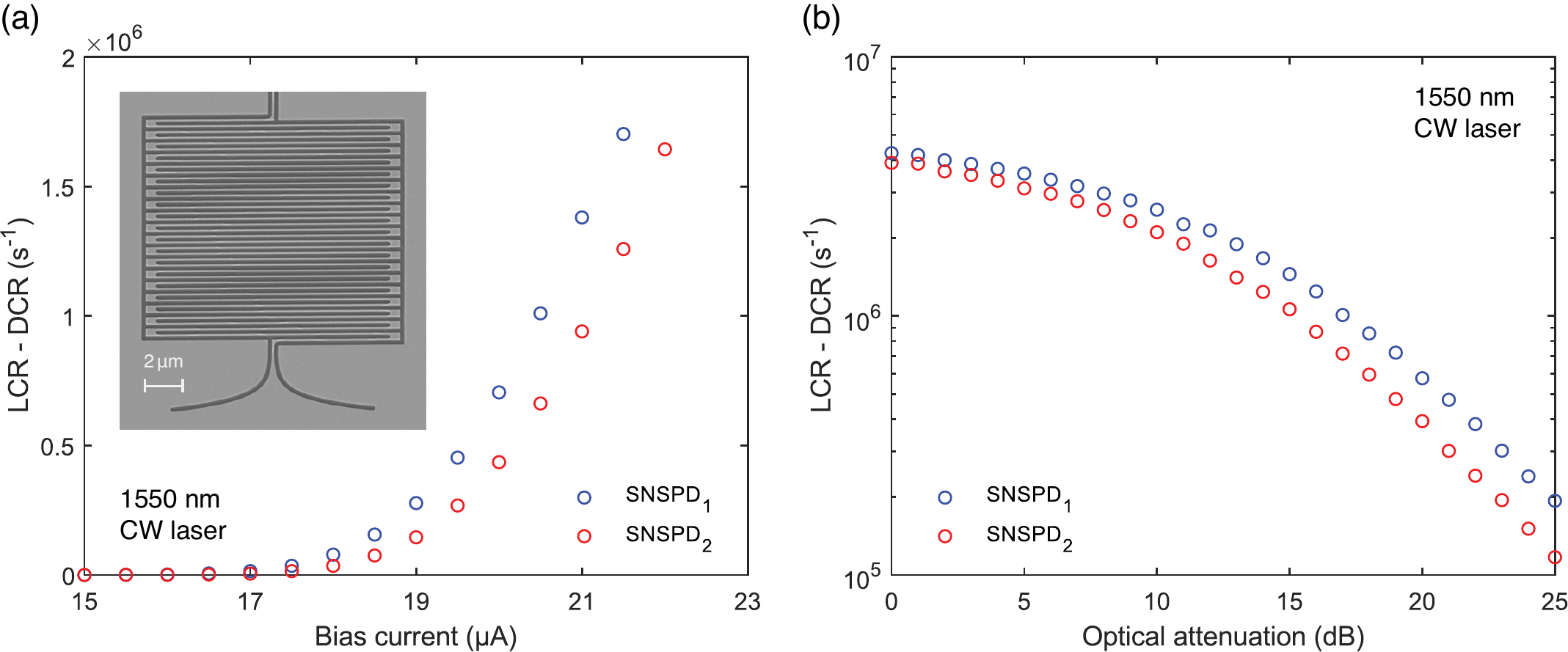}
    \caption{SNSPD characterization. (a) Light count rate (LCR) minus dark count rate (DCR) as a function of bias current for the two SNSPDs used to drive the logic gate. The detector was flood illuminated with a 1550\,nm continuous-wave (CW) laser. The optical attenuation was set to 15\,dB (0\,dB gives 50\,µW at the attenuator output). The inset shows the scanning electron micrograph of one detector. (b) Light count rate (LCR) minus dark count rate (DCR) as a function of optical attenuation for the two SNSPDs, both biased at 21\,µA. The single-photon detection regime starts above a 19\,dB attenuation for SNSPD$_1$, and 15\,dB for SNSPD$_2$ ($\alpha \approx 0.95$). $\alpha$ is the slope of the count rate in dB units versus attenuation.}
    \label{fig:SNSPD}
\end{figure}

The two detectors (SNSPD$_1$ and SNSPD$_2$) used to drive the logic gate are meandered 100\,nm wide wires, as shown in the inset of Fig.~\ref{fig:SNSPD}(a), fabricated on the same chip as the circuit. 
We characterized the performance of both detectors while they were already coupled to the circuit, and therefore shunted with surface-mount resistors. 
Fig.~\ref{fig:SNSPD}(a) shows their light count rate as a function of bias current, when flood illuminated with a 1550\,nm continuous-wave laser. 
As expected, given the film thickness and operating temperature, neither detector exhibits a saturation plateau. SNSPD$_1$ generates higher count rates for the same bias points, probably due to the presence of constrictions in the devices. 

Fig.~\ref{fig:SNSPD}(b) shows the count rate as a function of optical attenuation when the two detectors are biased at 21\,µA. As expected, also in this case the counts of SNSPD$_1$ are higher for the same attenuation. 
For both detectors, low attenuation values (below 19\,dB for SNSPD$_1$ and below 15\,dB for SNSPD$_2$) are associated with multi-photon detection events, while single-photon detection is confirmed at higher attenuations. 
Using the 1550\,nm pulsed laser, the curve would be slightly different but the general shape would be similar, with a multi-photon regime at higher powers. The experiment should be repeated at higher optical attenuation to experimentally confirm coincidence detection at the single-photon level. However, the circuit performance does not depend on the optical power, and the bit error rate is expected to remain consistent with the value obtained here.

\section{Bit-error rate and pulse-time analysis for the SNSPD-driven logic gate}

\begin{figure}[H]
    \centering
    \includegraphics[width=1\linewidth]{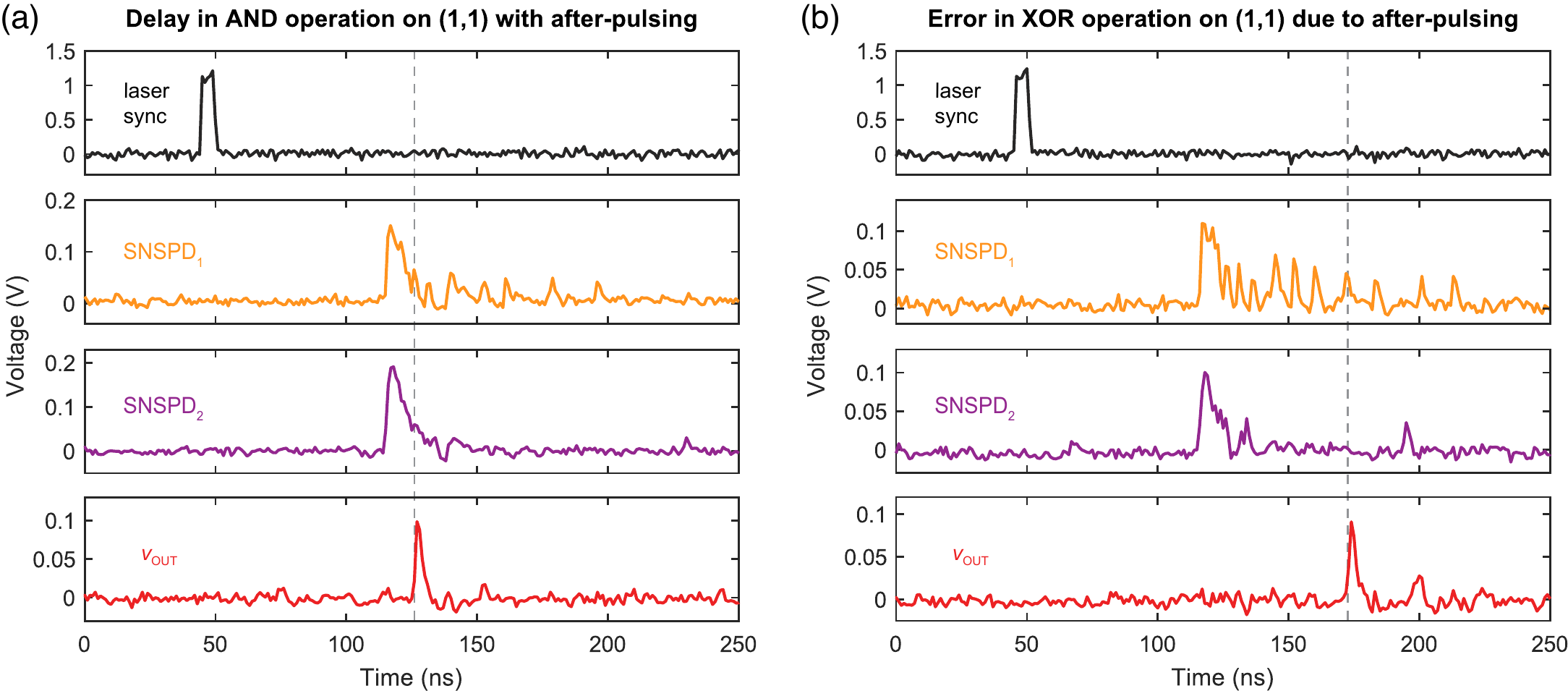}
    \caption{Examples of bit errors and delays on the output pulse of the SNSPD-driven gate. (a) Time-domain traces of a single AND operation on (1,1), with a 10\,ns delay between output and detector pulses ($I_\text{bs1,2} = 21$\,µA, $I_\text{bg1,2} = 14$\,µA, $I_\text{b1} = 83$\,µA, $I_\text{b2} = 87$\,µA, $I_\text{b3} = 110$\,µA). 
    (b) Time-domain traces of a single XOR operation on (1,1) with bit error after a 57\,ns delay from the detector pulses. ($I_\text{bs1} = 21$\,µA, $I_\text{bs2} = -21$\,µA, $I_\text{bg1} = 10$\,µA, $I_\text{bg2} = -4$\,µA, $I_\text{b1} = 94$\,µA, $I_\text{b2} = -106$\,µA, $I_\text{b3} = 105$\,µA). In each case, the panels show (from top to bottom) the 1\,MHz sync signal of the pulsed laser, the amplified output pulses of SNSPD$_1$ and SNSPD$_2$, and the amplified circuit output voltage. For the XOR operation, the output pulses of SNSPD$_2$ have negative amplitude, but the signal is inverted for clarity. The gray dashed lines highlight the time correspondence between output pulse and an afterpulse of SNSPD$_1$.}
    \label{fig:errors}
\end{figure}

During the BER estimation for XOR and AND operations on SNSPD outputs, we observed multiple types of errors that affected bit-error rate and timing resolution of the circuit: from delays in the output pulse arrival time to bit errors in single operations. 
In both configurations, we observed afterpulsing on the detector outputs, with a larger effect in XOR mode, due to higher bias currents and thus higher switching probabilities.  

Fig.~\ref{fig:errors}(a) shows a close-up view of the time-domain traces in the AND mode, where the output pulse is generated with an unintended delay of approximately 10\,ns relative to the SNSPD pulses. This delay is not systematic and occurs only with a certain probability.
We note that the output pulse is temporally aligned with a short afterpulse observed in the SNSPD$_1$ signal, suggesting that afterpulsing, reflection generation, and the signal reduction induced by the presence of the SNSPD readout line may be correlated with this effect.

Fig.~\ref{fig:errors}(b) shows a close-up view for the XOR operation on (1,1), where an error occurs: an unexpected output pulse is generated, with a 57\,ns delay relative to the SNSPD pulses. 
Also in this case, the output aligns with an afterpulse on the SNSPD$_1$ signal. Moreover, the number of afterpulses is much higher than in the AND mode, and there is a clear asymmetry in the afterpulsing of the two detectors.

\begin{figure}[H]
    \centering
    \includegraphics[width=1\linewidth]{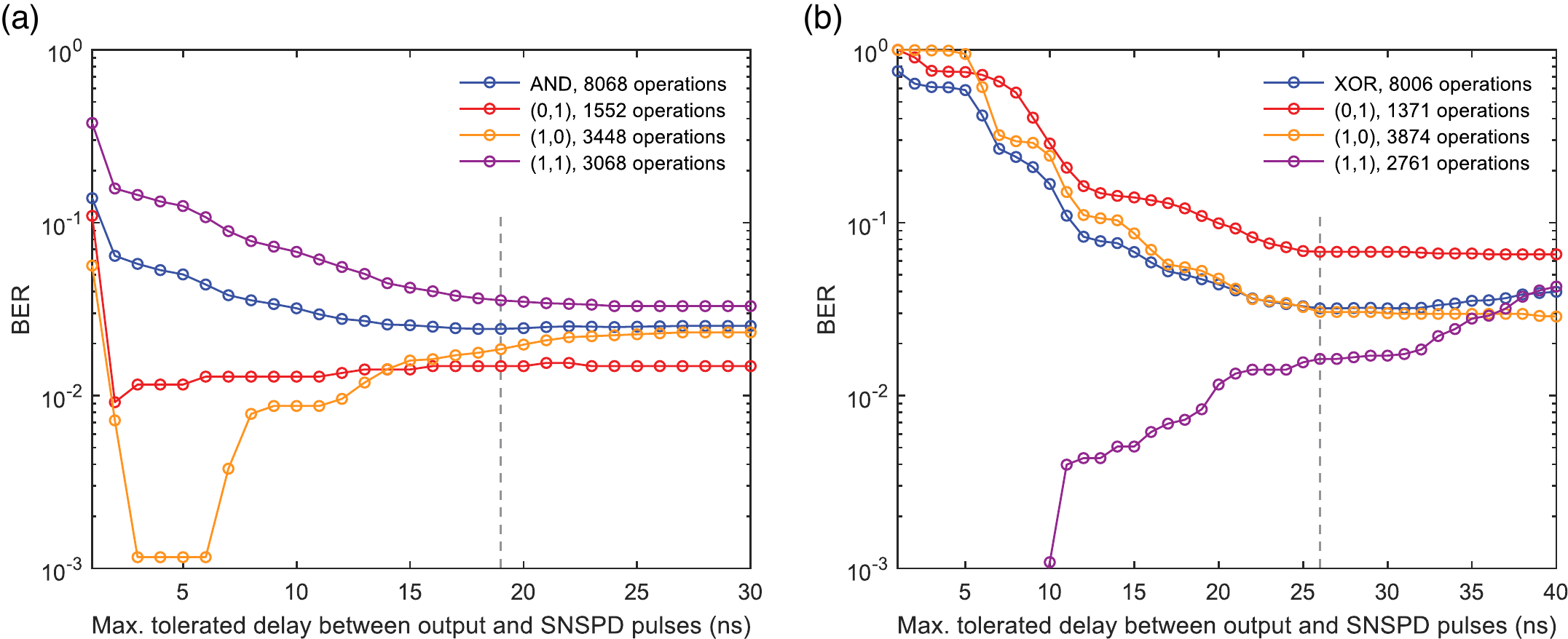}
    \caption{Bit-error rate of the SNSPD-driven gate for different operations as a function of the maximum tolerated delay between output and SNSPD pulses. (a) AND mode ($I_\text{bs1,2} = 21$\,µA, $I_\text{bg1,2} = 14$\,µA, $I_\text{b1} = 83$\,µA, $I_\text{b2} = 87$\,µA, $I_\text{b3} = 110$\,µA). (a) XOR mode ($I_\text{bs1} = 21$\,µA, $I_\text{bs2} = -21$\,µA, $I_\text{bg1} = 10$\,µA, $I_\text{bg2} = -4$\,µA, $I_\text{b1} = 94$\,µA, $I_\text{b2} = -106$\,µA, $I_\text{b3} = 105$\,µA). In both cases, red, yellow, and purple traces are associated with the BER for different single operations. The blue trace is the BER considering the three operations together. The gray dashed lines indicate the optimal delay for minimum BER on the blue trace: 19\,ns for AND; 26\,ns for XOR.}
    \label{fig:BER_delay}
\end{figure}

To take into account these types of errors and to study the effect of a variable delay in the output pulse, we estimated the BER under different timing constraints on the output. Specifically, we defined a time window $\tau$ following the arrival time of the SNSPD pulse, within which we checked whether an output pulse was present (output = 1) or absent (output = 0). We then calculated the BER over $10^4$ synchronized pulses for different values of the time window, corresponding to different maximum tolerated output delays.
Fig.~\ref{fig:BER_delay} shows how the BER for the AND and XOR operations varies as a function of the time window used for its extraction. It also includes the rates for each input combination to facilitate error analysis.

In AND mode, the BER varies from 13\% at $\tau = 1$\,ns to 2.5\% for $\tau \ge 22$\,ns, with a minimum of 2.43\% at $\tau = 19$\,ns. 
The error associated with the (1,1) input combination dominates the BER. The strong dependence of this BER on the time window $\tau$ for $\tau < 22$\,ns indicates that the output pulse is generated with a variable delay. In particular, the presence of plateaus in the BER curve confirm that in some cases the output can be triggered by SNSPD afterpulses, as illustrated in Fig.~\ref{fig:errors}. Indeed, the width of these plateaus is comparable to the period of the afterpulses. The plateau observed for $\tau \ge 22$\,ns indicates that there is also a finite probability that the output does not fire at all. 
For the (0,1) and (1,0) input combinations, the BER is generally lower. In these cases, afterpulses can cause the output to fire when it should not, with a higher probability for SNSPD$_1$, consistent with its stronger afterpulsing behavior.

In XOR mode, the BER varies from 75\% at $\tau = 1$\,ns to 3.98\% for $\tau \ge 40$\,ns, with a minimum value of 3.19\% at $\tau = 26$\,ns. In this mode, the dependence on the time window is significantly stronger, and the dominant error occurs for the (0,1) input combination. This error is primarily associated with the output firing due to afterpulses in SNSPD$_1$ for $\tau < 34$\,ns, or not firing at all with a probability of 6.56\%.
The error for the (1,0) input combination is generally lower because SNSPD$_1$ exhibits stronger afterpulsing and therefore a higher probability of triggering an output pulse. The errors for the (1,1) input combination, illustrated in Fig.~\ref{fig:errors}(b), are very low or absent for short $\tau$ and increase for larger $\tau$, as a wider time window allows a higher probability of unwanted output pulses generated by afterpulsing. 

In a practical application, the dependence of the BER on the output delay could be problematic; therefore, the circuit should be optimized to mitigate these effects. Fabricating SNSPDs coupled to the logic gate with on-chip resistors would surely improve control on the electrical coupling.  With the current circuit, a possible solution would be to operate it in a clocked configuration to limit the acceptable timing uncertainty of the output pulses. 

\section{Energy consumption analysis in LTspice}

To quantitatively evaluate the energy efficiency of the logic gate operating on two SNSPDs' outputs, we performed transient simulations in LTspice using the same bias conditions as in the experiment (Results in Fig.~4 and setup in Fig.~\ref{fig:setup_snspd}). The objective was to extract the energy per operation for XOR and AND functions across all input combinations.

The experimental DC voltage sources with series bias resistors ($R_{\text{b}} = 10\,\text{k}\Omega$) were approximated as ideal DC current sources with infinite output impedance in LTspice. This approximation was valid because all circuit elements were shunted by resistances in the range of $5$--$50\,\Omega \ll 10\,\text{k}\Omega$. The energy dissipated by the bias resistors was excluded from the calculation.

Using the exact bias conditions from the experiment and the nominal values of gate widths, inductors, and resistors, the gate of nT$_3$ did not switch when expected. We therefore reduced the choke width of nT$_3$ from 45\,nm (used for nT$_1$ and nT$_2$) to 33\,nm to obtain a switching event.

To extract the energy per operation, we monitored the voltage waveforms of each current source supplying current $I_i$, and we computed the instantaneous power: $p_i(t) = v_i(t)I_i$.
We then integrated $p_i(t)$ over the active pulse window of a single logic operation to obtain the total energy delivered by each source. The window was defined from a time preceding the photon arrival to the time at which all currents in the circuit had reset back into the superconducting elements.
The sum of all source contributions, including those used to bias the SNSPDs, yielded the total dissipated energy for a given logic event, shown in Table~1.

To isolate the energy consumption of the circuit itself, we calculated the power dissipated by the SNSPDs and subtracted it from the total. The instantaneous power dissipated by each SNSPD was computed by recording the hotspot resistance and applying the relation: $p(t) = R_\text{hs}(t)i^2(t)$. The energy values are shown in Table~1.

Each SNSPD's pulse generated afterpulses caused by oscillations in the choke of the side nTrons, a phenomenon consistent with the afterpulses observed experimentally. In simulations, each afterpulse dissipated approximately 1\,aJ, with 5 to 24 afterpulses observed per SNSPD pulse on each side nTron, depending on the bias conditions and logic operation. These contributions are included in the total energy values reported in Table~1. For each operation, we also report the consumption of the circuit itself, assuming no afterpulses are generated.

\begin{table}[htbp]
    \centering
    
    \label{tab:mylabel}
    \begin{tabular}{|l|c|c|c|}
\hline
         & $E_{(1,0)}$ (aJ) & $E_{(0,1)}$ (aJ) & $E_{(1,1)}$ (aJ) \\
        \hline
        AND + SNSPDs      & 165 & 183 & 447 \\
        AND      & 113 & 134 & 344 \\
        AND w/o afterpulses      & 89 & 109 & 298 \\
        \hline
        XOR + SNSPDs       & 244 & 230 & 316 \\
        XOR     & 193 & 188 & 216 \\
        XOR w/o afterpulses       & 175 & 183 & 193 \\
\hline
    \end{tabular}
    \caption{Estimated energy consumption per operation for different logic functions, SNSPD output combinations, and energy contributions (with or without detectors, with or without afterpulses).}
\end{table}

\section{nTron driving a capacitor}

\begin{figure}[H]
    \centering
    \includegraphics[width=0.98\linewidth]{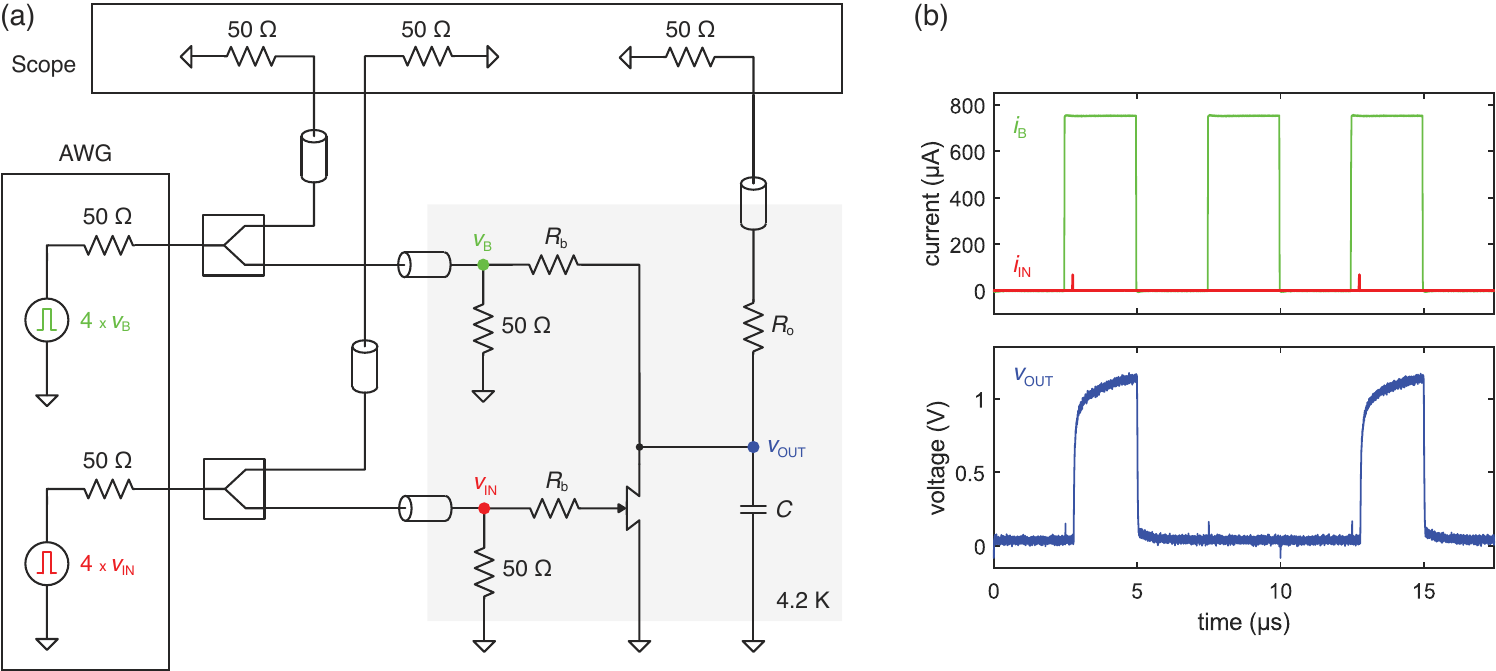}
    \caption[nTron driving a capacitive load.]{Driving a capacitor with an nTron. (a) Experimental setup: the input and clock signals, generated by an arbitrary waveform generator (sources 1 and 2 of the AWG), are split and read by the oscilloscope, as well as being sent to the device ($v_\text{IN}$ and $v_\text{B}$); coaxial cables connect the device from 4.2\,K to room-temperature electronics; 50\,$\Omega$ resistors in front of the bias resistors ($R_\text{b} = 5.11\,\text{k}\Omega$) ensure matching between the PCB and the coaxial cable (cylindrical elements) for both input and clock signals; the components enclosed in the gray box are on the PCB at 4.2\,K; the load capacitor $C$ is wire-bonded to the nTron's drain; the output signal passes through a $10\,\text{k}\Omega$ resistor ($R_\text{o}$), before going to the oscilloscope to ensure a high load impedance for the nTron; the impedance of the scope is set to 50\,$\Omega$.  
    (b) Experimental time-domain response: the top panel shows the clocked bias current ($i_\mathrm{B} = v_\text{B}/R_\text{b}$), and the input signal ($i_\mathrm{IN} = v_\text{IN}/R_\text{b}$); the amplitudes of the input and clock pulses are 68\,\textmu A and 753\,\textmu A respectively; the clock frequency is 200\,kHz; the bottom panel shows the capacitor voltage ($v_\mathrm{OUT}$), obtained by multiplying the voltage on the scope by the factor $(50\,\Omega + R_\text{o})/50\,\Omega$; currents and voltages are averaged across 100 traces.
    }
    \label{fig:nT_capa}
\end{figure}

We verified that the nTron can drive a capacitive load to voltages compatible with electro-optic modulator control levels ($\sim1$\,V), demonstrating its suitability for direct modulator integration. A capacitor was chosen as the load because, at the tested frequencies ($\sim~100$\,kHz), the input impedance of electro-optic modulators, whether traveling-wave or ring resonator based, is well approximated by a lumped capacitance.
Fig.~\ref{fig:nT_capa}(a) shows the schematic of the setup used for this verification. 
The circuit consists of a single nTron (1.5\,\textmu m channel, 200\,nm choke, 10\,nm thickness) biased close to its critical current, with a 1\,pF capacitor wire bonded to its drain. 
We applied input pulses of 5\,ns duration and 68\,\textmu A amplitude to the nTron gate at 100\,kHz ($i_\text{IN}$), in phase with a clocked bias current ($i_\text{B}$) supplied to the channel at 200\,kHz to periodically reset the device. The clocked bias (753\,\textmu A amplitude) is necessary because the high load impedance and low kinetic inductance of the drain prevent self-resetting. We set the clock frequency to twice the input frequency to ensure the nTron does not switch on bias pulses alone, in the absence of an input pulse.

\vspace{0.2cm}

To maximize the output voltage, we did not connect the 50\,$\Omega$ readout line directly to the capacitor; instead, we added a 10\,k$\Omega$ series resistor ($R_\text{o}$) to prevent heavy shunting of the nTron channel, allowing the hotspot resistance to grow sufficiently large. We recovered the capacitor voltage from the scope trace by multiplying by the factor $(50\,\Omega + R_\text{o})/50\,\Omega$, and averaged the signal over 100 traces.

Fig.~\ref{fig:nT_capa}(b) shows the time-domain traces of the input current, clock signal, and output voltage. For each clock period coinciding with an input pulse, the capacitor voltage exhibits a fast initial rise to 0.6\,V in 30\,ns ($\sim0.02$\,V/ns), followed by a slow saturation to 1.15\,V over 2.2\,\textmu s. The device resets rapidly when the clock current is removed, with a fall time of 33\,ns (measured from 90\% to 10\% of the amplitude). The slow charging dynamics are governed by a combination of capacitive charging and hotspot growth, which produce a time-dependent impedance and hence a non-constant current delivered to the capacitor. Indeed, assuming a constant current $I_\text{B}$ charging the capacitor, the expected rise time would be $t = CV_\text{max}/I_\text{B} = 1.5$\,ns, far shorter than the experimentally observed 2.2\,\textmu s. Similarly, the fall time is influenced by the finite slew rate of the clock signal at its falling edge (4\,ns width). A more detailed analysis of these dynamics through LTspice simulations would help analyze the problem in the future.
The small 120\,mV transient pulses visible on $v_\text{OUT}$ are generated by the fast rising and falling edges of the clock signal.

Operating at higher repetition rates than 100\,kHz caused the device to switch on clock pulses alone, even in the absence of an input pulse, suggesting incomplete thermal relaxation between cycles. By reducing the clock bias current, and thus the resistive heating, stable operation at higher repetition rates was recovered, at the cost of a reduced output voltage. Future efforts should focus on optimizing thermal dissipation to mitigate this trade-off.